\documentclass[
    aps,
    prd,
    10pt,
    twocolumn,
    nofootinbib,
    notitlepage,
    superscriptaddress,
    showkeys,
    floatfix
]{revtex4-2}

\usepackage{amsmath,amssymb}
\usepackage{mathtools}
\usepackage{bm}
\usepackage{braket}
\usepackage{physics}
\usepackage{slashed}
\usepackage{mathrsfs}

\usepackage{graphicx}
\usepackage{dcolumn}
\usepackage{multirow}

\usepackage{xcolor}
\usepackage{soul}
\usepackage{tikz}

\usepackage[mathlines]{lineno}

\usepackage[
    colorlinks=true,
    citecolor=blue,
    linkcolor=blue,
    urlcolor=blue
]{hyperref}

\definecolor{lime}{HTML}{A6CE39}
\DeclareRobustCommand{\orcidicon}{%
	\begin{tikzpicture}
	\draw[lime, fill=lime] (0,0) 
	circle [radius=0.16] 
	node[white] {{\fontfamily{qag}\selectfont \tiny ID}};
	\draw[white, fill=white] (-0.0625,0.095) 
	circle [radius=0.007];
	\end{tikzpicture}
	\hspace{-2mm}
}
\foreach \x in {A, ..., Z}{%
	\expandafter\xdef\csname orcid\x\endcsname{\noexpand\href{https://orcid.org/\csname orcidauthor\x\endcsname}{\noexpand\orcidicon}}
}

\begin{document}
\title{Skewness dependence of the pion and kaon generalized parton distributions}

\author{Fernando Chandra\orcidB{}}
\email[E-mail: ]{fernando.chandra@ui.ac.id} 
\affiliation{Departemen Fisika, FMIPA, Universitas Indonesia, Depok 16424, Indonesia}

\author{Parada T.~P.~Hutauruk\orcidA{}}
\email[E-mail: ]{phutauruk@hiroshima-u.ac.jp}
\affiliation{International Institute for Sustainability with Knotted Chiral Meta Matter (WPI-SKCM$^2$), Hiroshima University, Higashi-Hiroshima, Hiroshima 739-8526, Japan}

\author{Terry Mart\orcidC{}}
\email[E-mail: ]{terry.mart@sci.ui.ac.id}
\affiliation{Departemen Fisika, FMIPA, Universitas Indonesia, Depok 16424, Indonesia}

\date{\today}

\begin{abstract}
We investigate the skewness dependence of the pion generalized parton distribution (GPD) at different momentum transfers $-t$ within the covariant Nambu-Jona-Lasinio (NJL) model, employing the Schwinger proper-time regularization scheme to regulate ultraviolet divergences and implement an effective description of quark confinement. We evolve the pion GPDs for various values of $-t$ and $\xi$ to the scales $\mu^2=4$ and $27~\mathrm{GeV}^2$. We find that the valence-quark distributions are suppressed with increasing $-t$ and $\xi$, with the suppression becoming more pronounced at the lower scale, $\mu^2=4~\mathrm{GeV}^2$. In the forward limit, $\xi=0$ and $-t=0$, the predicted pion valence-quark distribution at $\mu^2=27~\mathrm{GeV}^2$ is in good agreement with the available experimental data and the global JAM analysis. The corresponding pion gluon distribution at $\mu^2=4~\mathrm{GeV}^2$, obtained through next-to-leading-order (NLO) DGLAP evolution, is also consistent with the JAM analysis. We further find that the pion and kaon valence-quark distributions exhibit a strong dependence on $-t$ but only a weak dependence on $\xi$, whereas the corresponding gluon distributions show weak dependence on both $-t$ and $\xi$. The generalized form factors of the pion and kaon at $\mu^2=4~\mathrm{GeV}^2$ are in good agreement with the corresponding lattice-QCD results.
\end{abstract}
\keywords{Nambu--Jona-Lasinio model, generalized parton distributions, pseudo-Goldstone boson, generalized form factor.}

\maketitle

\section{Introduction}
\label{sec:intro}
The pion and kaon, as the lightest bound states of QCD and pseudo-Goldstone bosons, provide a crucial avenue for exploring the underlying dynamics of the strong interaction, particularly its nonperturbative features, including dynamical chiral symmetry breaking (DCSB)~\cite{Zweig:1964ruk,Zweig:1964jf,Lee:1972fj}, asymptotic freedom~\cite{Gross:2022hyw,Shibata:2018tyx}, and quark confinement. The internal structure of hadrons is encoded in generalized parton distributions (GPDs)~\cite{Diehl:2003ny,Belitsky:2005qn,Goeke:2001tz,Ji:1998pc,Radyushkin:1997ki}, which provide a unified description of the information contained in parton distribution functions and electromagnetic form factors. GPDs thus provide simultaneous access to the longitudinal momentum fraction and transverse spatial distributions of partons within a hadron. Moreover, they provide valuable information on the electromagnetic and mechanical structure of hadrons.

On the theoretical side, pion and kaon GPDs at zero skewness, $\xi=0$, have been studied using a variety of models with different approaches, including the light-cone wave-function (LCWF) overlap representation~\cite{Vogt:2000ku,Vogt:2001if}, the light-cone quark model (LCQM)~\cite{Luan:2024dvc}, the Bethe-Salpeter equation-Nambu-Jona-Lasinio (BSE-NJL) model~\cite{Chandra:2025pqs,Zhang:2021shm,Zhang:2021tnr,Zhang:2025iiw}, the nonlocal chiral quark model (NLChQM)~\cite{Son:2024uet}, and holographic QCD or the AdS/QCD correspondence~\cite{Kaur:2018ewq}. Global QCD analyses have also been employed to investigate pion GPDs at zero skewness~\cite{Goharipour:2025zsw,Guo:2022upw}.

More recently, pion and kaon GPDs at nonzero skewness have been investigated in a number of theoretical frameworks~\cite{Son:2024uet,Kaur:2018ewq,Bakulev:2000eb,Polyakov:1999gs,Mankiewicz:1997uy,Anikin:2000th,Choi:2001fc,Choi:2002ic,Mukherjee:2002gb,Tiburzi:2002kr,Tiburzi:2002tq,Theussl:2002xp,Broniowski:2003rp,Bissey:2003yr,Broniowski:2007si,Vogt:2001if,Hoodbhoy:2003uu,Chavez:2021llq,Guo:2025muf}. These studies have attracted considerable interest because pion and kaon GPDs provide access to the three-dimensional spatial structure (tomography) of these mesons~\cite{Burkardt:2000za} and are directly related to the energy-momentum tensor (EMT) and the decomposition of hadron mass~\cite{Ji:1996ek}. In parallel, recent lattice-QCD calculations have been used to determine the odd Mellin moments of the pion valence-quark GPD at both zero and nonzero skewness~\cite{Lin:2023gxz,Gao:2025inf,Chen:2019lcm,Lin:2020rxa,Guo:2025muf}.

On the experimental side, pion and kaon GPDs can, in principle, be accessed through hard exclusive processes~\cite{Boer:2025ixc}, such as deeply virtual meson production (DVMP)~\cite{Amrath:2008vx} and the Sullivan process~\cite{Sullivan:1971kd}. However, experimental information on pion and kaon GPDs remains scarce, primarily because free pion and kaon targets are not readily available. Consequently, these distributions remain poorly constrained. In the near future, more precise measurements of pion and kaon structure are expected from modern facilities, including the Electron-Ion Collider (EIC)~\cite{Arrington:2021biu}, the Electron-ion Collider in China (EicC)~\cite{Anderle:2021wcy}, the Apparatus for Meson and Baryon Experimental Research (AMBER)/COMPASS++ at CERN~\cite{Adams:2018pwt}, the J-PARC Hadron Experimental Facility Extension Project~\cite{Sakuma:2022twx,Aoki:2021cqa}, and the 22-GeV upgrade of Jefferson Lab (JLab)~\cite{Accardi:2023chb,Mart:2026tyg}.

In this paper, we investigate pion and kaon GPDs at nonzero skewness within the covariant Nambu-Jona-Lasinio (NJL) model, employing the proper-time regularization scheme to regulate the divergent one-loop quark momentum integrals and implement an effective description of quark confinement. The NJL model has been successfully applied to a wide range of hadronic and nuclear phenomena, including the valence-quark distributions of the pion and kaon~\cite{Hutauruk:2016sug,Braghin:2026hkt,Hutauruk:2018zfk}, their electromagnetic form factors~\cite{Ninomiya:2014kja}, meson fragmentation functions~\cite{Matevosyan:2010hh}, the EMC and polarized EMC effects~\cite{Cloet:2006bq}, nuclear matter~\cite{Lawley:2006ps}, and meson properties at finite temperature and density~\cite{Hutauruk:2021dgv}. This work extends our previous study of the pion GPD at zero skewness presented in Ref.~\cite{Chandra:2025pqs}. We investigate the pion and kaon GPDs for momentum transfers $-t=0$, $0.11$, $0.5$, $1.0$, and $1.5~\mathrm{GeV}^2$ and skewness values $\xi=0.05$, $0.15$, and $0.25$. The resulting distributions are evolved to the higher renormalization scales $\mu^2=4$ and $27~\mathrm{GeV}^2$ using the next-to-leading-order (NLO) Dokshitzer-Gribov-Lipatov-Altarelli-Parisi (DGLAP) evolution equations~\cite{Miyama:1995bd}. We compare our results with the available experimental data~\cite{E615:1989bda}, other theoretical calculations, lattice-QCD results, and the global JAM analysis~\cite{Barry:2021osv}.

The remainder of this paper is organized as follows. In Sec.~\ref{sec:njl}, we briefly review the covariant Nambu-Jona-Lasinio (NJL) model with the Schwinger proper-time regularization scheme, which provides an effective implementation of quark confinement and is used to calculate the quark and meson properties. In Sec.~\ref{sec:gpd}, we present the expressions for the pion and kaon GPDs within the NJL model, beginning with the generic leading-twist-2 GPD expressions and discussing their Mellin moments and associated polynomiality properties. In Sec.~\ref{sec:num}, we present our numerical results for the pion and kaon GPDs and their Mellin moments for $-t=0$, $0.11$, $0.5$, $1.0$, and $1.5~\mathrm{GeV}^2$ and $\xi=0$, $0.05$, $0.15$, and $0.25$ at the scales $\mu^2=4$ and $27~\mathrm{GeV}^2$, obtained by evolving from the initial scale $\mu_0^2=0.18~\mathrm{GeV}^2$. Finally, we summarize our findings and present our conclusions in Sec.~\ref{sec:sum}.

\section{SU(3) NJL model}
\label{sec:njl}
Here, we describe the pion and kaon properties and internal structure in the covariant SU(3) flavor NJL model Lagrangian. The expression for the SU(3) flavor NJL model can be defined in terms of the four-fermion interactions as follows~\cite{Hutauruk:2025gnb,Gifari:2024ssz,Klevansky:1992qe}
\begin{eqnarray}
    \label{eq:njl1}
    \mathscr{L}_{\mathrm{NJL}} &=& \bar{\psi}_q \big(i\partial\!\!\!/-\hat{m}_q \big) \psi_q \nonumber \\
    &+& G_\pi \big[ (\bar{\psi}_q \lambda_a \psi_q)^2-(\bar{\psi}_q \lambda_a \gamma_5 \psi_q)^2\big] \nonumber \\
    &-& G_V \big[ (\bar{\psi}_q \lambda_a \gamma^\mu  \psi_q)^2 + (\bar{\psi}_q \lambda_a \gamma^\mu \gamma_5 \psi_q)^2\big],
\end{eqnarray}
where $\psi_q = (\psi_u, \psi_d, \psi_s)^T$ is the quark field with flavor $q= (u,d,s)$, $\hat{m}_q =\mathrm{diag} (m_u,m_d,m_s)$ stands for the current (bare) quark matrix. The $\lambda_a$ with $a=1, \cdot \cdot \cdot, 8$ are the Gell-Mann matrices in flavor space with $\lambda_0  \equiv \sqrt{\frac{1}{3}} \mathbf{1}$. Note that the $\eta$ and $\eta^\prime$ mesons and six-fermion interaction~\cite{Klevansky:1992qe} are usually considered in the Lagrangian in Eq.~(\ref{eq:njl1}), which explicitly breaks the global U(1) symmetry. However, in this work, we do not consider such terms for simplicity, and the terms will not directly influence our results on pion and kaon properties. The coupling constant $G_\pi$ represents the four-fermion interaction term in scalar and pseudoscalar meson channels, which is responsible for the dynamical breaking of chiral symmetry (DBCS) and generating the dressed quark masses. The coupling $G_V$ represents the quark-antiquark interaction in the vector and axial-vector meson channels. 

The NJL gap equation for each quark flavor $q =(u, d,s)$ in the proper-time regularization scheme can be written as
\begin{eqnarray}
    M_q = m_q + \frac{3M_q G_\pi}{\pi^2} \int \frac{d\tau}{\tau^2} \, \exp[-\tau(M_q^2)].
\end{eqnarray}

In the NJL model, pseudoscalar mesons are realized as bound states of a dressed quark and a dressed antiquark, whose properties can be determined by solving the Bethe-Salpeter equation (BSE). The BSE solutions for the pion and kaon channels can be obtained following the procedure described in Ref.~\cite{Hutauruk:2016sug}. Consequently, the reduced $t$-matrices in the pseudoscalar-meson channels can be written as
\begin{eqnarray}
    t_{M} (q) &=& \frac{-2iG_\pi}{1+ 2 G_\pi \Pi_{M}(q^2)},
\end{eqnarray}
with the bubble (polarization insertion) diagram,
\begin{eqnarray}
    \Pi_{M} (q^2) = 2iN_c \int \frac{d^4k}{(2\pi)^4} \mathrm{Tr}_D \big[\gamma_5 S_q (k) \gamma_5 S_{q'} (k+q)\big].
\end{eqnarray}

The pseudoscalar meson masses are determined from the pole condition of the corresponding $t$-matrix, which is given by
\begin{eqnarray}
    1 + 2 G_\pi \Pi_{M} (k^2 = m_{M}^2) = 0.
\end{eqnarray}
Analytically, the pseudoscalar meson masses in the proper time regularization scheme can be written as
\begin{eqnarray}
\label{eq:mesonmass}
    m_M^2 =\Big[\frac{m_{q'}}{M_{q'}}+ \frac{m_q}{M_q} \Big] \frac{1}{G_\pi I_{qq'}(m_{M}^2)} + \big( M_{q'} -M_q \big)^2,
\end{eqnarray}
where $I_{qq'} (k^2) = (N_C/\pi^2)\int^1_0 dx \int \frac{d\tau}{\tau} \exp \big[ -\tau \big(x(x-1)k^2 +xM_q^2 +(1-x)M_{q'}^2 \big)\big]$. The analytic meson mass expression in Eq.~(\ref{eq:mesonmass}) clearly demonstrates the pseudo-Goldstone boson nature of the pion and kaon in the chiral limit. Rather than the meson masses, we can also obtain the meson-quark coupling constants for the pseudoscalar mesons from the residue at the pole of the $q\bar{q}$ $t$-matrices. This gives
\begin{eqnarray}
    Z_{M}^{-1} = (g_{Mqq})^{-2} = - \frac{\partial \Pi_{M} (q^2)}{\partial q^2} \Bigg|_{q^2 = m_{M}^2}.
\end{eqnarray}

The parameters determined within the NJL model, such as $M_q$, $m_{M}$, and $g_{Mqq}$, are used as inputs for calculating the pion and kaon GPDs at nonzero skewness, as described in detail in Sec.~\ref{sec:gpd}.

\section{Kaon and pion GPDs}
\label{sec:gpd}
Here, we briefly introduce the vector GPDs for the pseudoscalar mesons within the framework of the covariant NJL model. However, before we present the expression of pseudoscalar meson GPDs in the NJL model, we start from the generic expression of GPDs, which is formulated in the nonlocal quark-quark matrix element in the nondiagonal momentum space and the quark field separated along the light-cone. It is defined as 
\begin{equation}
 \label{eqgpd1a}
\mathcal{H}^q (x,\xi,t) =  \int_{z} e^{\left( ixP^+ z^- \right)} \big< {M} (p') \mid J^+_q \mid {M}(p) \big>_{z^+=0,\mathbf{z}=0},
\end{equation}
where $\int_{z} =\int {dz^-}/{4\pi}$, $J^+_q = \bar{\psi}_q \big(-{z^-}/{2} \big) \gamma^+ \psi_q \big( {z^-}/{2} \big)$ with $\psi_q$ being the quark field with flavor $q= (u,d,s)$, $x$ is the Bjorken variable or the quark longitudinal momentum fraction, $t = Q^2 = -q^2 = \Delta^2 = (p'-p)^2$ being the momentum transfer, and $\xi = (p^+ -p'^+)/(p^+ + p'^+)= -\Delta^+/2p^+$ is the skewness parameter. The average momentum of the meson is given by $P = (p +p')/2$ with $p$ and $p'$ being the initial and final pion momentum, respectively. The pseudoscalar mesons can be categorized by the leading-twist vector (no spin flip) $\mathcal{H}^q(x,\xi,t)$ and tensor (spin flip) $E^q(x,\xi,t)$ quark GPDs. Note that, in the present work, we only focus on the spin-non-flip vector pion and kaon GPDs. The relation between the quark GPDs and the isoscalar and isovector GPDs can be given by
\begin{eqnarray}
    \mathcal{H}^{I=0} (x,\xi,t) &=& \mathcal{H}^q (x,\xi,t) + \mathcal{H}^{\bar{q}} (x,\xi,t), \\
    \mathcal{H}^{I=1} (x,\xi,t) &=& \mathcal{H}^q (x,\xi,t) - \mathcal{H}^{\bar{q}} (x,\xi,t),
\end{eqnarray}
where $\mathcal{H}^q (x,\xi,t)$ supports $x \in [0,1]$ and $\mathcal{H}^{\bar{q}} (x,\xi,t) = - \mathcal{H}^q (-x,\xi,t)$ supports $x \in [-1+ \xi,\xi]$. The $x \in [0,\xi]$ is the so-called Efremov-Radyushkin-Brodsky-Lepage (ERBL) region, and $x \in [\xi,1]$ is called the DGLAP region. It is worth noting that, in the forward limit, the meson GPDs for $t=0$ and $\xi=0$ will simply reduce to the meson PDFs, where in this kinematic limit, the initial and final meson momenta are equal $p =p'$, and it gives
\begin{eqnarray}
    \mathcal{H}^q_M (x, 0,0) &=& \frac{1}{2} \big[ \mathcal{H}^{I=0} (x,0,0) + \mathcal{H}^{I=1} (x,0,0) \big], \nonumber \\
    &=& q_v^{M} (x),
\end{eqnarray}
where $ q_v^{M} (x) = q_{M} (x) - \bar{q}_{M} (x)$ is the valence quark distribution of the meson with $M =(\pi,K)$ being the pseudoscalar meson type. The valence quark distribution of the meson should satisfy the number and momentum sum rules,
\begin{eqnarray}
    1&=&\int_0^1 dx\,\, q_v^{M} (x), \\
    1 &=&\int_0^1 dx \,\,x[q_{M}(x) +\bar{q}_{M} (x)  + \bar{q}'_{M} (x) + q'_{M}(x) ] .~~~~
\end{eqnarray}

Based on the polynomiality condition, the meson GPDs can be related to the generalized form factors through their Mellin moments: 
\begin{eqnarray}
    \int^1_{-1} dx\,\, x^n \mathcal{H}^{q}_M (x,\xi,t) &=& \sum_{i=0}^{l} \xi^{2i} A_{n+1,2i}^{qM} (t), 
\end{eqnarray}
where $l=\left\lfloor (n+1)/2 \right\rfloor$, with $\left\lfloor \cdots \right\rfloor$ denoting the floor function, and $A_{n+1,2i}^{qM}(t)$ are the generalized form factors for arbitrary values of $n$ and $i$. In particular, for $n=0$, the zeroth moment gives the quark vector form factor, 
\begin{eqnarray}
    \int^1_{-1} dx\,\, \mathcal{H}^q_M (x,\xi=0,t) &=& A_{10}^{qM} (t) = F_{M}^q (Q^2).
\end{eqnarray}
\begin{figure}[t]
\centering
\includegraphics[width=1\columnwidth]{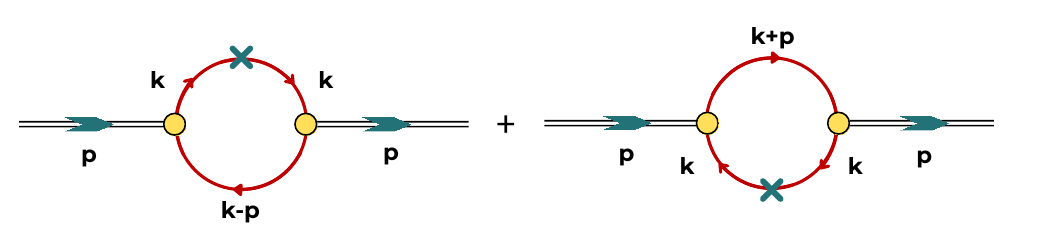} 
\caption{\label{figfd} Two dominant Feynman diagrams contribute to the pseudoscalar meson GPDs in the NJL model. } 
\end{figure}

In the NJL model, the dominant contribution to meson GPDs originates from the two Feynman diagrams shown in Fig.~\ref{figfd}. Accordingly, the meson GPDs can be expressed mathematically as
\begin{eqnarray}
    \mathcal{H}^q_M (x,\xi,t) &=& 2iN_C g_{Mqq}^2 \int \frac{d^4k}{(2\pi)^4} \delta \big( xP^+ -k^+ \big)\nonumber \\
    &\times& \mathrm{Tr} \big[\gamma_5 S_q (k) \gamma^+ S_{q} (k) \gamma_5 S_{q'} (k-P)],~~~
\end{eqnarray}
where $S_q (k) = (k\!\!\!/ + M_q)/(k^2 -M_q^2 + i \epsilon)$ is the quark propagator with the flavor $q =(u,d,s)$.

After lengthy analytical calculations, the resulting expressions for the pseudoscalar meson GPDs in the NJL model, using the proper-time regularization scheme, can be written as
\begin{eqnarray}
    \label{eq:gpdnjl}
    \mathcal{H}^q_M (x,\xi,t) &=& \frac{N_C g_{M qq}^2}{8\pi^2} \int_{\tau_{\mathrm{UV}}^2}^{\tau_{\mathrm{IR}}^2} \frac{d\tau}{\tau} \Theta_{01} \exp \big[ -\tau \big( \Delta_1 \big) \big] \nonumber \\
    &+& \frac{N_C g_{M qq}^2}{8\pi^2} \int_{\tau_{\mathrm{UV}}^2}^{\tau_{\mathrm{IR}}^2} \frac{d\tau}{\tau} \Theta_{23} \exp\big[ -\tau \big( \Delta_2 \big)  \big] \nonumber \\
    &+& \frac{N_C xg_{{M} qq}^2 }{8\pi^2 \xi} \int_{\tau_{\mathrm{UV}}^2}^{\tau_{\mathrm{IR}}^2} \frac{d\tau}{\tau} \Theta_{45} \exp\big[ -\tau \big( \Delta_0 \big) \big] \nonumber \\
    &+& \frac{N_C g_{{M} qq}^2}{16\pi^2\xi}  \Theta (\alpha_6) \Theta (\alpha_7) \big[(1-x)t +  B \big] \nonumber \\
    &\times& \int_{\tau_{\mathrm{UV}}^2}^{\tau_{\mathrm{IR}}^2} d\tau \int_0^{1}d{\beta}  \,\Theta\left(1-\beta-\beta_1\right) \nonumber \\
    &\times&\exp\big[ -\tau \big( \Delta_3 \big) \big], 
\end{eqnarray}
where the subscript $M$ stands for the meson type, i.e., $M =$ $K$ for kaon and $M=$ $\pi$ for pion. However, it is worth noting that the GPD expression in Eq.~(\ref{eq:gpdnjl}) is a generic expression for different quark masses, which is appropriate for the kaon. For the pion, it can be straightforwardly determined by replacing $M_q \rightarrow M_{q'}$  or $M_s \rightarrow M_u $. Other variable quantities are defined as 
\begin{eqnarray}
    \Theta_{01} &=& \Theta (\alpha_{0} ) \Theta ( \alpha_1 ), \\
    \Theta_{23} &=& \Theta (\alpha_2 )  \Theta ( \alpha_3),\\ \Theta_{45} &=& \Theta (\alpha_4)  \Theta (\alpha_5),\\ \Delta_1 &=& M_{q'}^2 - \alpha _1 (M_{q'}^2-M_q^2)-\alpha_1 (1-\alpha_1) m_{M}^2, \\
    \Delta_2 &=& M_{q'}^2 - \alpha _2 (M_{q'}^2-M_q^2)-\alpha_2 (1-\alpha_2) m_{{M}}^2,~~ \\
    \Delta_0 &=& M_q^2 - \beta_0 (1-\beta_0) t,\\
    \Delta_3 &=& M_q^2-\beta(M_q^2-M_{q'}^2) - {\beta} (1-{\beta})m_{{M}}^2 \nonumber \\
    &-&\beta_1 (1-\beta_1 -{\beta}) t, \\
    B &= & 2x\big[m_{{M}}^2- (M_{q}-M_{q'})^2\big].
\end{eqnarray}
Furthermore, the variables in Eq.~(\ref{eq:gpdnjl}) are given by
\begin{align*}
    \alpha_0 &= \frac{x+\xi}{1+\xi}, & \alpha_1 &= \frac{1-x}{1+\xi}, \\
    \alpha_2 &= \frac{x-1}{\xi-1}, & \alpha_3 &= \frac{\xi-x}{\xi-1}, \\
    \alpha_4 &= 1-\frac{x}{\xi}, & \alpha_5 &= 1+\frac{x}{\xi}, \\
    \alpha_6 &= \frac{\xi+x-(1+\xi)\beta}{\xi}, & \alpha_7 &= \frac{\xi-x+(1-\xi)\beta}{\xi}, \\
    \beta_0 &= \frac{1}{2}\alpha_5, & \beta_1 &= \frac{1}{2}\alpha_7.
\end{align*}
Here we emphasize again that the pseudoscalar meson GPDs in Eq.~(\ref{eq:gpdnjl}) provide a generic formulation applicable to arbitrary quark masses. Since we focus on the kaon and pion GPDs in this work, we evaluate these GPDs at nonzero skewness at $\mu^2 =$ 4 and 27 GeV$^2$, with their scale dependence evolved using the NLO-DGLAP evolution equations~\cite{Miyama:1995bd}.

\section{Numerical Result}
\label{sec:num}
Numerical results for the valence-quark distribution of the pion and kaon for different values of $-t$ and $\xi$ at $\mu^2 =$ 4 and 27 GeV$^2$, as well as their Mellin moments, are presented in Figs.~\ref{fig1}-\ref{fig4}. The NJL parameter set used in this calculation is determined by taking $M_u=M_d=0.4~\mathrm{GeV}$, assuming SU(2) isospin symmetry, together with $\Lambda_{\rm IR}=0.24~\mathrm{GeV}$, $m_\pi=0.14~\mathrm{GeV}$, $f_\pi=0.093~\mathrm{GeV}$, $m_K=0.495~\mathrm{GeV}$, and $f_K=0.097~\mathrm{GeV}$. These input parameters yield $G_\pi=19.04~\mathrm{GeV}^{-2}$, $\Lambda_{\rm UV}=0.645~\mathrm{GeV}$, $g_{\pi qq}=4.225$, $M_s=0.611~\mathrm{GeV}$, and $g_{Kqq}=4.570$, which are adapted from Refs.~\cite{Hutauruk:2016sug,Hutauruk:2021dgv}.

\subsection{Results for pion GPDs}
\begin{figure}[t]
\centering
\includegraphics[width=1\columnwidth]{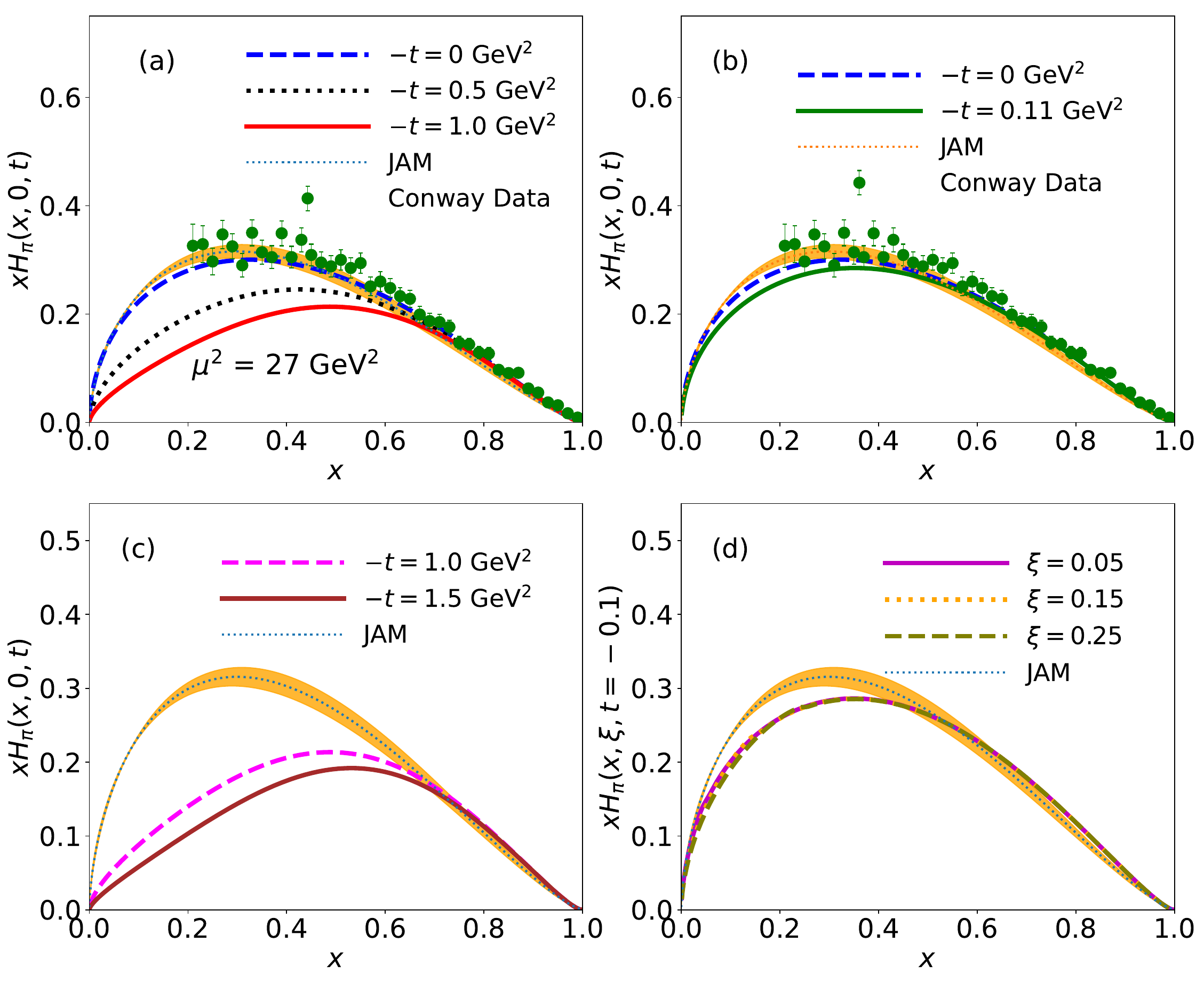} 
\caption{Valence-quark distributions of the pion at the scale $\mu^2=27~\mathrm{GeV}^2$ for (a) $-t=0$, $0.5$, and $1.0~\mathrm{GeV}^2$, (b) $-t=0$ and $0.11~\mathrm{GeV}^2$, (c) $-t=2$ and $5~\mathrm{GeV}^2$, and (d) $\xi=0$, $0.05$, $0.15$, and $0.25$, shown as functions of $x$ and compared with the available experimental data~\cite{E615:1989bda} and the JAM QCD analysis~\cite{Barry:2021osv}.}
\label{fig1}
\end{figure}

Figure~\ref{fig1}(a) shows the pion valence-quark distribution, $xH_\pi^u(x,0,t)$, for $-t=0$, $0.5$, and $1.0~\mathrm{GeV}^2$ at $\xi=0$ and $\mu^2=27~\mathrm{GeV}^2$, together with the available experimental data~\cite{E615:1989bda} and the JAM analysis~\cite{Barry:2021osv}. In the forward limit, $-t=0$ and $\xi=0$, our result for $xH_\pi^u(x,0,0)$ is in good agreement with the experimental data~\cite{E615:1989bda} and the JAM analysis~\cite{Barry:2021osv}, consistent with the results reported in Refs.~\cite{Hutauruk:2016sug,Chandra:2025pqs}. For $-t=0.5$ and $1.0~\mathrm{GeV}^2$, we find that the valence-quark distribution is increasingly suppressed as $-t$ increases. This behavior is consistent with other theoretical calculations~\cite{Son:2024uet,Kaur:2018ewq} and lattice-QCD results~\cite{Guo:2025muf}.

For fixed $-t=1.5~\mathrm{GeV}^2$ and $\xi=0$ at $\mu^2=27~\mathrm{GeV}^2$, the first Mellin moment of the pion valence-quark distribution is $\langle x\rangle_v^\pi=0.119$, compared with $\langle x\rangle_v^\pi=0.200$ in the forward limit. This substantial reduction demonstrates the suppression of the quark momentum fraction with increasing momentum transfer. A detailed comparison of the Mellin moments for different values of $-t$ and $\xi$ is provided in Table~\ref{tab4}.
\begin{table*}[t]
	\begin{ruledtabular}
		\renewcommand{\arraystretch}{1.2}
		\caption{Mellin moments of the pion valence-quark distributions for $-t=0$, $0.11$, $0.5$, $1.0$, and $1.5~\mathrm{GeV}^2$ and $\xi=0.05$, $0.15$, and $0.25$ at the scales $\mu^2=4$ and $27~\mathrm{GeV}^2$.}
		\label{tab4}
		\begin{tabular}{c|c|c|c|ccccccc}
		 $\big< x^n \big>_{v}^{\pi}$  & \boldmath{$\mu^2$ (GeV$^2$)} & \boldmath{$\xi$} & \boldmath{$-t$ (GeV$^2$)} & \boldmath{$n=1$} & \boldmath{$n=2$}  & \boldmath{$n=3$} & \boldmath{$n=4$}  & \boldmath{$n=5$} \\[1ex] \hline\\[-1.5ex]
             & $\mathbf{4}$  & 0.0 & 0.0 & 0.230 & 0.103 & 0.059 & 0.038 & 0.026 \\  
              &  & 0.0 & 0.11 & 0.219 & 0.100 & 0.057 & 0.037 & 0.026 \\  
            &  & 0.0& 0.5 & 0.187 & 0.091 & 0.054 & 0.035 & 0.025\\  
             &  &0.0& 1.0 & 0.158 & 0.081 & 0.050 & 0.033 & 0.024 \\
            &  & 0.0 & 1.5 & 0.137 & 0.074 & 0.046 & 0.032 & 0.023\\  
           \hline\\[-1.5ex]
            &  &0.05& 0.1 & 0.220 & 0.100 & 0.057 & 0.037 & 0.026 \\ 
            &  &0.15& 0.1 & 0.219 & 0.100 & 0.057 & 0.037 & 0.026 \\ 
            &  &0.25& 0.1 & 0.218 & 0.100 & 0.057 & 0.037  & 0.026 \\ \hline\\[-1.5ex]
             &  &  &  &  &   &  &  \\ \hline \hline \\[-1.5ex]
            & $\mathbf{27}$  & 0.0 & 0.0 & 0.200 & 0.083 & 0.045 & 0.028 & 0.019\\  
              &  & 0.0 & 0.11 & 0.191 & 0.081 & 0.044 & 0.027 & 0.018\\  
            &  & 0.0& 0.5 & 0.163 & 0.073 & 0.041 & 0.026  & 0.018\\  
            &  &0.0& 1.0 & 0.138 & 0.066 & 0.038 & 0.025 & 0.017 \\  
            &  & 0.0 & 1.5 & 0.119 & 0.060 & 0.036 & 0.023 & 0.016\\  
           \hline\\[-1.5ex]
            &  &0.05& 0.1 & 0.191 & 0.081 & 0.044 & 0.027 & 0.018\\ 
            &  &0.15& 0.1 & 0.191 & 0.081 & 0.044 & 0.027 & 0.018\\ 
            &  &0.25& 0.1 & 0.190 & 0.081 & 0.044 & 0.027  & 0.018 
		\end{tabular}
	\end{ruledtabular}
\end{table*}

In Fig.~\ref{fig1}(b), we show the pion valence-quark distribution at fixed $-t=0.11~\mathrm{GeV}^2$ in comparison with the forward-limit result at $-t=0$. We find that the distribution at $-t=0.11~\mathrm{GeV}^2$ differs only slightly from that in the forward limit. Nevertheless, the valence-quark distribution decreases with increasing $-t$.

At the larger momentum transfer, $-t=1.5~\mathrm{GeV}^2$, the pion valence-quark distribution is further suppressed, as clearly shown in Fig.~\ref{fig1}(c). This behavior is consistent with the expected $t$ dependence of the distribution.

In addition to varying $-t$, we also compute the pion valence-quark distribution for different values of the skewness, $\xi$, at $\mu^2=27~\mathrm{GeV}^2$, as shown in Fig.~\ref{fig1}(d). We find that the pion valence-quark distribution decreases with increasing $\xi$, particularly in the region $0\leq x\leq0.6$. For $x\gtrsim0.6$, the distribution exhibits the opposite behavior, increasing with increasing $\xi$. Our results are consistent with other theoretical calculations~\cite{Son:2024uet,Kaur:2018ewq} and the recent lattice-QCD results~\cite{Gao:2025inf}.
\begin{figure}[t]
\centering
\includegraphics[width=1\columnwidth]{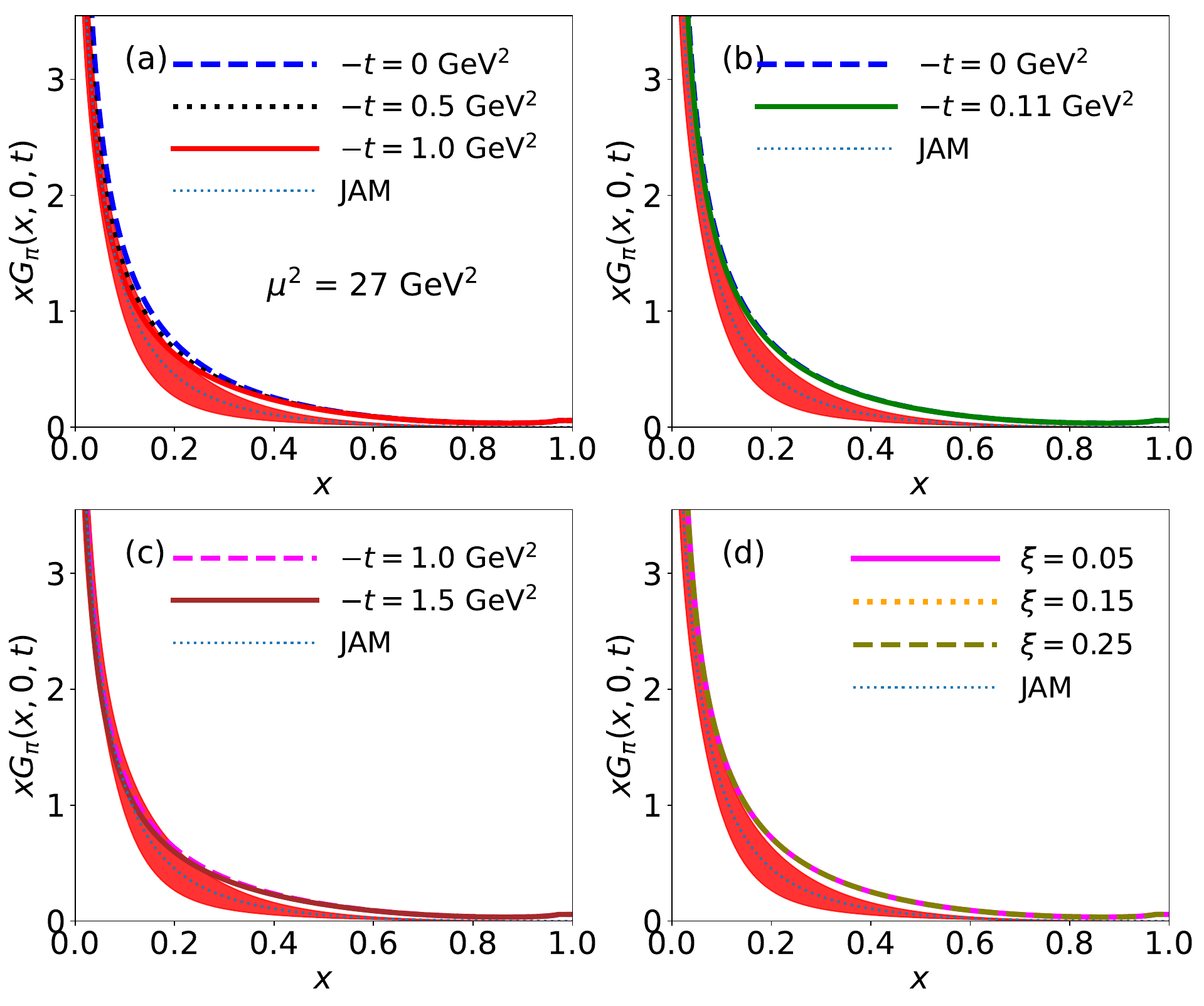} 
\caption{Gluon distribution of the pion at the scale $\mu^2 =$ 27 GeV$^2$ in comparison with the JAM QCD analysis with the same scale~\cite{Barry:2021osv}. } 
\label{fig2}
\end{figure}

We further calculate the pion gluon distributions for the corresponding values of $-t$, shown in Fig.~\ref{fig1}, obtained by evolving the pion valence-quark distributions using the NLO DGLAP evolution equations~\cite{Miyama:1995bd}. Figure~\ref{fig2}(a) shows the pion gluon distributions at fixed $\xi=0$ for $-t=0$, $0.5$, and $1.0~\mathrm{GeV}^2$. We find that the gluon distribution exhibits only a weak dependence on $-t$ and decreases slightly as $-t$ increases, similar to the behavior of the valence-quark distribution. This finding is consistent with Ref.~\cite{Kaur:2025gyr}, where the pion gluon distribution was found to decrease with increasing $-t$ and $\xi$. In the forward limit, our pion gluon distribution is in good agreement with the corresponding JAM analysis~\cite{Barry:2021osv}. A similar behavior is observed in Fig.~\ref{fig2}(b) for the smaller values of $-t$ considered there. We further consider the larger momentum transfers, $-t=1.0$ and $1.5~\mathrm{GeV}^2$, and find that the pion gluon distribution remains relatively insensitive to $-t$, with only minor changes in magnitude compared with the forward-limit result at $-t=0$ and $\xi=0$.

Similar to the pion valence-quark distribution, we compute the pion gluon distribution at fixed $-t=0.1~\mathrm{GeV}^2$ for $\xi=0.05$, $0.15$, and $0.25$, as shown in Fig.~\ref{fig2}(d). We find that the pion gluon distribution exhibits only a weak dependence on $\xi$, with its magnitude changing only slightly as $\xi$ increases. Nevertheless, the distribution decreases gradually with increasing $\xi$. 
\begin{figure}[t]
\centering
\includegraphics[width=1\columnwidth]{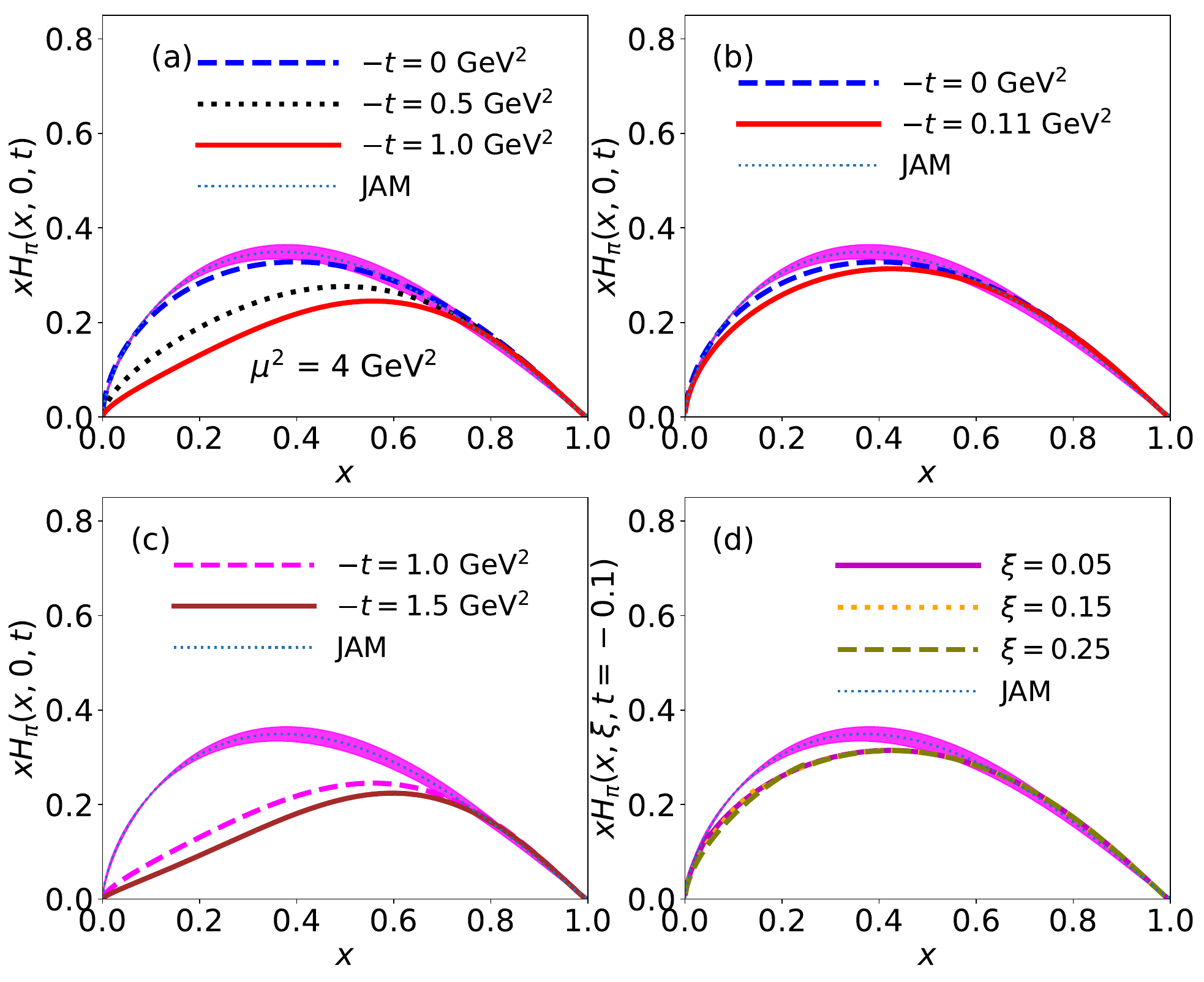} 
\caption{Same as in Fig.~\ref{fig1}, but for the renormalization scale $\mu^2 = 4$ GeV$^2$ in comparison with the JAM QCD analysis with the same scale~\cite{Barry:2021osv}.}
\label{fig3} 
\end{figure}

The pion valence-quark distributions for different values of $-t$ and $\xi$ at the scale $\mu^2=4~\mathrm{GeV}^2$ are shown in Figs.~\ref{fig3}(a)-\ref{fig3}(d). Figure~\ref{fig3}(a) shows that the valence-quark distribution decreases with increasing $-t$. The distributions also exhibit differences in magnitude compared with the corresponding results shown in Fig.~\ref{fig1}(a). In the forward limit, $-t=0$ and $\xi=0$, our pion valence-quark distribution at $\mu^2=4~\mathrm{GeV}^2$ is consistent with the corresponding JAM analysis~\cite{Barry:2021osv}, in agreement with the results reported in Ref.~\cite{Chandra:2025pqs}. The first Mellin moment of the pion valence-quark distribution at fixed $-t=1.5~\mathrm{GeV}^2$ and $\xi=0$ is found to be $\langle x\rangle_v^\pi=0.137$ at $\mu^2=4~\mathrm{GeV}^2$. The higher Mellin moments of the pion valence-quark distribution at the same scale are listed in Table~\ref{tab4}.

We also consider the smaller momentum transfer, $-t=0.11~\mathrm{GeV}^2$, at $\xi=0$, as shown in Fig.~\ref{fig3}(b). We find that the pion valence-quark distribution exhibits only a small change in magnitude compared with the forward-limit result at $-t=0$ and $\xi =0$. Nevertheless, the distribution decreases with increasing $-t$.

We subsequently consider the larger momentum transfer, $-t=1.5~\mathrm{GeV}^2$, as shown in Fig.~\ref{fig3}(c). We find that the pion valence-quark distribution at $\mu^2=4~\mathrm{GeV}^2$ is more strongly suppressed at $-t=1.5~\mathrm{GeV}^2$ than at $-t=0.5$ and $1.0~\mathrm{GeV}^2$.

Furthermore, Fig.~\ref{fig3}(d) shows the pion valence-quark distributions for $\xi=0.05$, $0.15$, and $0.25$ at fixed $-t=0.1~\mathrm{GeV}^2$. We find that the pion valence-quark distributions decrease with increasing $\xi$, particularly in the region $0\leq x\leq0.6$. At larger values of $x$, the distributions for different values of $\xi$ tend to overlap and are consistent with the JAM analysis~\cite{Barry:2021osv}. The first and higher Mellin moments of the pion valence-quark distributions for different values of $\xi$ and $-t$ at the scale $\mu^2=4~\mathrm{GeV}^2$ are listed in Table~\ref{tab4}.
\begin{figure}[t]
\centering
\includegraphics[width=1\columnwidth]{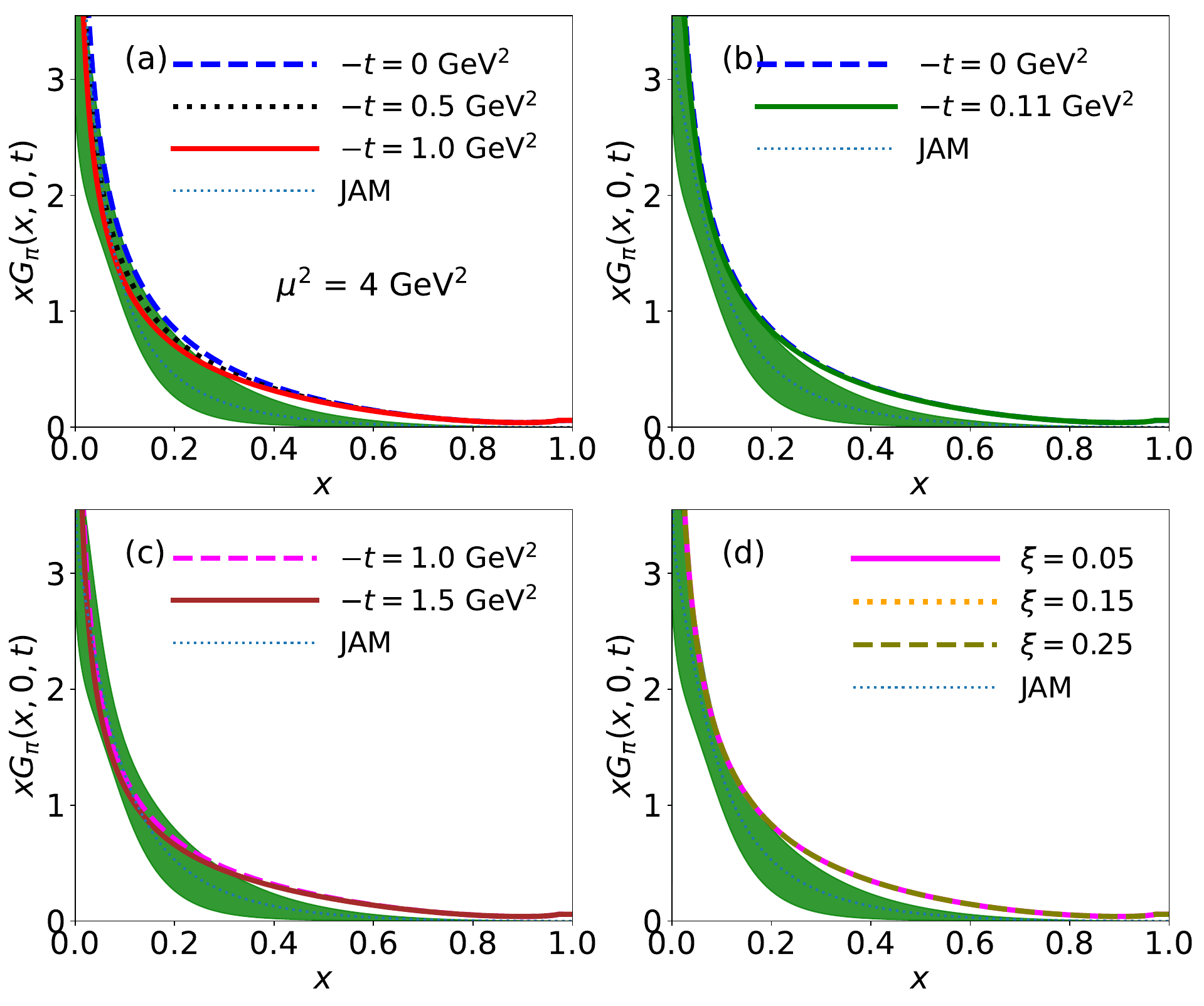} 
\caption{\label{fig4} Same as in Fig.~\ref{fig2}, but for the scale $\mu^2 = 4$ GeV$^2$ in comparison with the JAM QCD analysis with the same scale~\cite{Barry:2021osv}. } 
\end{figure}

Next, we present our results for the pion gluon distributions at different values of $\xi$ and $-t$ at the scale $\mu^2=4~\mathrm{GeV}^2$, as shown in Figs.~\ref{fig4}(a)-\ref{fig4}(d). For $\xi=0$ and $-t=0$, $0.5$, and $1.0~\mathrm{GeV}^2$, the corresponding pion gluon distributions are shown in Fig.~\ref{fig4}(a). We observe that the gluon distribution decreases with increasing $-t$. In the forward limit, $\xi=0$ and $-t=0$, our result at $\mu^2=4~\mathrm{GeV}^2$ is in good agreement with the corresponding JAM analysis~\cite{Barry:2021osv}. Similar behavior is observed in Figs.~\ref{fig4}(b)-\ref{fig4}(d), where the pion gluon distribution decreases with increasing $\xi$ and/or $-t$. However, the dependence on $\xi$ and $-t$ is relatively weak, resulting in only minor differences among the distributions for the different kinematic configurations considered.

\subsection{Results for kaon GPDs}
\begin{table*}[t]
	\begin{ruledtabular}
		\renewcommand{\arraystretch}{1.2}
		\caption{Mellin moments of the kaon up valence-quark distributions for $-t=0$, $0.11$, $0.5$, $1.0$, and $1.5~\mathrm{GeV}^2$ and $\xi=0.05$, $0.15$, and $0.25$ at the scales $\mu^2=4$ and $27~\mathrm{GeV}^2$.}
		\label{tab5}
		\begin{tabular}{c|c|c|c|ccccccc}
		 $\big< x^n \big>_{v}^{uK}$  & \boldmath{$\mu^2$ (GeV$^2$)} & \boldmath{$\xi$} & \boldmath{$-t$ (GeV$^2$)} & \boldmath{$n=1$} & \boldmath{$n=2$}  & \boldmath{$n=3$} & \boldmath{$n=4$}  & \boldmath{$n=5$} \\[1ex] \hline\\[-1.5ex]
             & $\mathbf{4}$  & 0.0 & 0.0 & 0.196 & 0.078 & 0.041 & 0.024 & 0.016 \\  
              &  & 0.0 & 0.11 & 0.185 & 0.075 & 0.040 & 0.024 & 0.016 \\  
            &  & 0.0& 0.5 & 0.154 & 0.067 & 0.037 & 0.023 & 0.015\\  
            &  & 0.0 & 1.0 & 0.126 & 0.059 & 0.034 & 0.021 & 0.015\\  
             &  &0.0& 1.5 & 0.107 & 0.053 & 0.031 & 0.020 & 0.014 \\
           \hline\\[-1.5ex]
            &  &0.05& 0.1 & 0.186 & 0.076 & 0.040 & 0.024 & 0.016 \\ 
            &  &0.15& 0.1 & 0.187 & 0.076 & 0.040 & 0.024 & 0.016 \\ 
            &  &0.25& 0.1 & 0.188 & 0.077 & 0.040 & 0.025  & 0.016 \\ \hline\\[-1.5ex]
             &  &  &  &  &   &  &  \\ \hline \hline \\[-1.5ex]
            & $\mathbf{27}$  & 0.0 & 0.0 & 0.170 & 0.063 & 0.031 & 0.018 & 0.012\\  
              &  & 0.0 & 0.11 & 0.161 & 0.061 & 0.030 & 0.018 & 0.011\\  
            &  & 0.0& 0.5 & 0.134 & 0.054 & 0.028 & 0.017  & 0.011\\  
             &  & 0.0 & 1.0 & 0.110 & 0.048 & 0.026 & 0.016 & 0.010\\  
            &  &0.0& 1.5 & 0.093 & 0.043 & 0.024 & 0.015 & 0.010 \\  
           \hline\\[-1.5ex]
            &  &0.05& 0.1 & 0.162 & 0.061 & 0.030 & 0.018 & 0.011\\ 
            &  &0.15& 0.1 & 0.163 & 0.062 & 0.031 & 0.018 & 0.011\\ 
            &  &0.25& 0.1 & 0.164 & 0.062 & 0.031 & 0.018  & 0.012 
		\end{tabular}
	\end{ruledtabular}
\end{table*}
\begin{figure}[t]
\centering
\includegraphics[width=1\columnwidth]{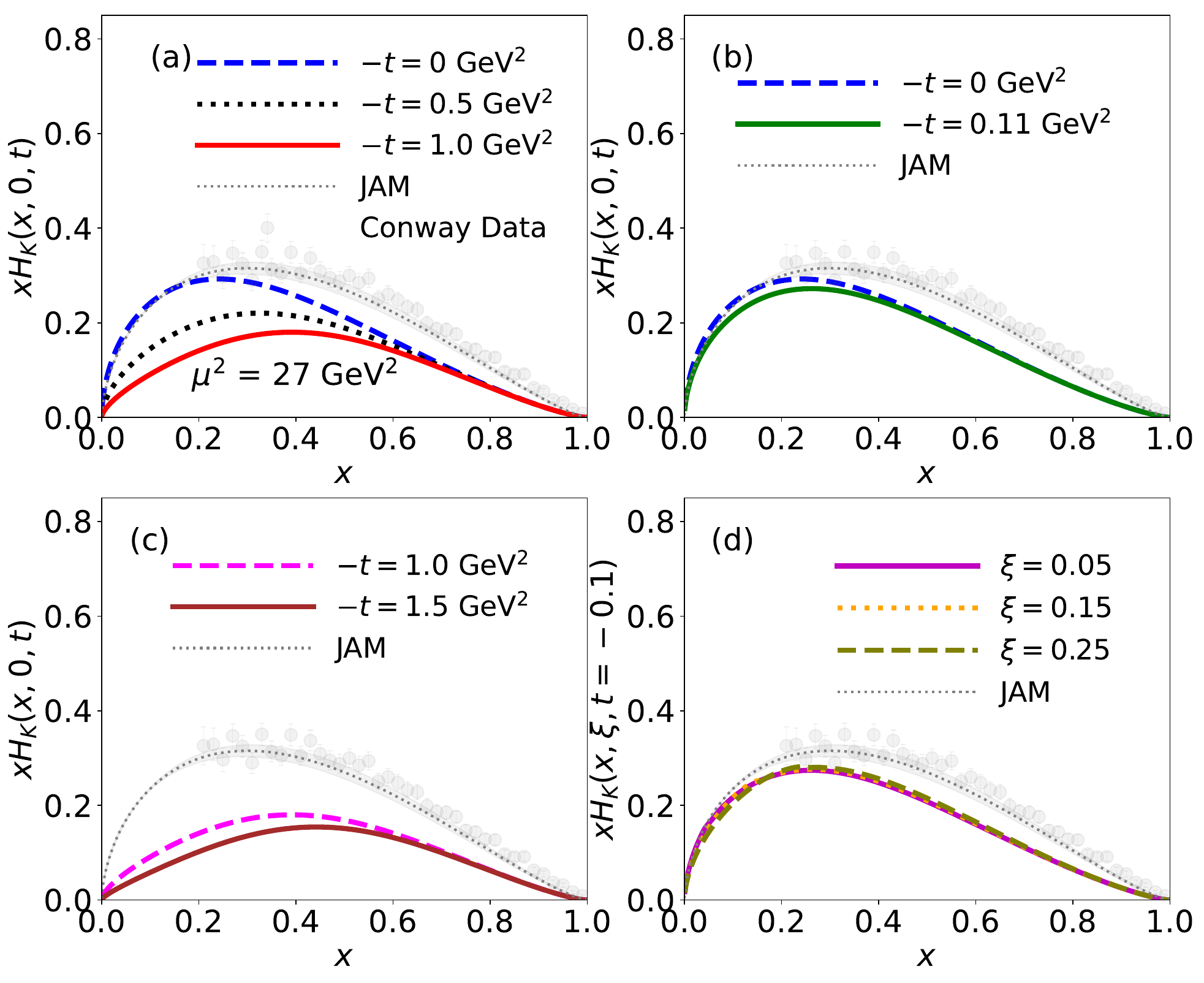} 
\caption{\label{fig5} Kaon valence-quark distributions at the scale $\mu^2=27~\mathrm{GeV}^2$ as functions of the longitudinal momentum fraction $x$ for (a) $-t=0$, $0.5$, and $1.0~\mathrm{GeV}^2$, (b) $-t=0$ and $0.11~\mathrm{GeV}^2$, (c) $-t=1.0$ and $1.5~\mathrm{GeV}^2$, and (d) $\xi=0$, $0.05$, $0.15$, and $0.25$. For reference, the pion valence-quark distribution from Ref.~\cite{E615:1989bda} and the JAM QCD analysis from Ref.~\cite{Barry:2021osv} are also shown.} 
\end{figure}

We now turn to the kaon valence-quark distribution at the scale $\mu^2=27~\mathrm{GeV}^2$, shown in Figs.~\ref{fig5}(a)-\ref{fig5}(d). Figure~\ref{fig5}(a) presents the kaon valence-quark distribution at fixed $\xi=0$ for $-t=0$, $0.5$, and $1.0~\mathrm{GeV}^2$. Similar to the pion case, the kaon valence-quark distribution is increasingly suppressed with increasing $-t$ in the region $0\leq x\leq0.6$. For $x\gtrsim0.6$, the distribution exhibits only a weak dependence on $-t$ and remains nearly unchanged. A more pronounced suppression is observed at $-t=1.5~\mathrm{GeV}^2$ compared with the forward-limit result, as shown in Fig.~\ref{fig5}(c).

In Fig.~\ref{fig5}(d), we present the kaon valence-quark distributions as functions of the longitudinal momentum fraction $x$ at fixed $-t=0.1~\mathrm{GeV}^2$ for $\xi=0.05$, $0.15$, and $0.25$. We find that the kaon valence-quark distribution exhibits only a weak dependence on $\xi$. In particular, the distribution decreases with increasing $\xi$ in the region $0\leq x\leq0.2$, whereas it increases with increasing $\xi$ for $x>0.2$.
\begin{figure}[t]
\centering
\includegraphics[width=1\columnwidth]{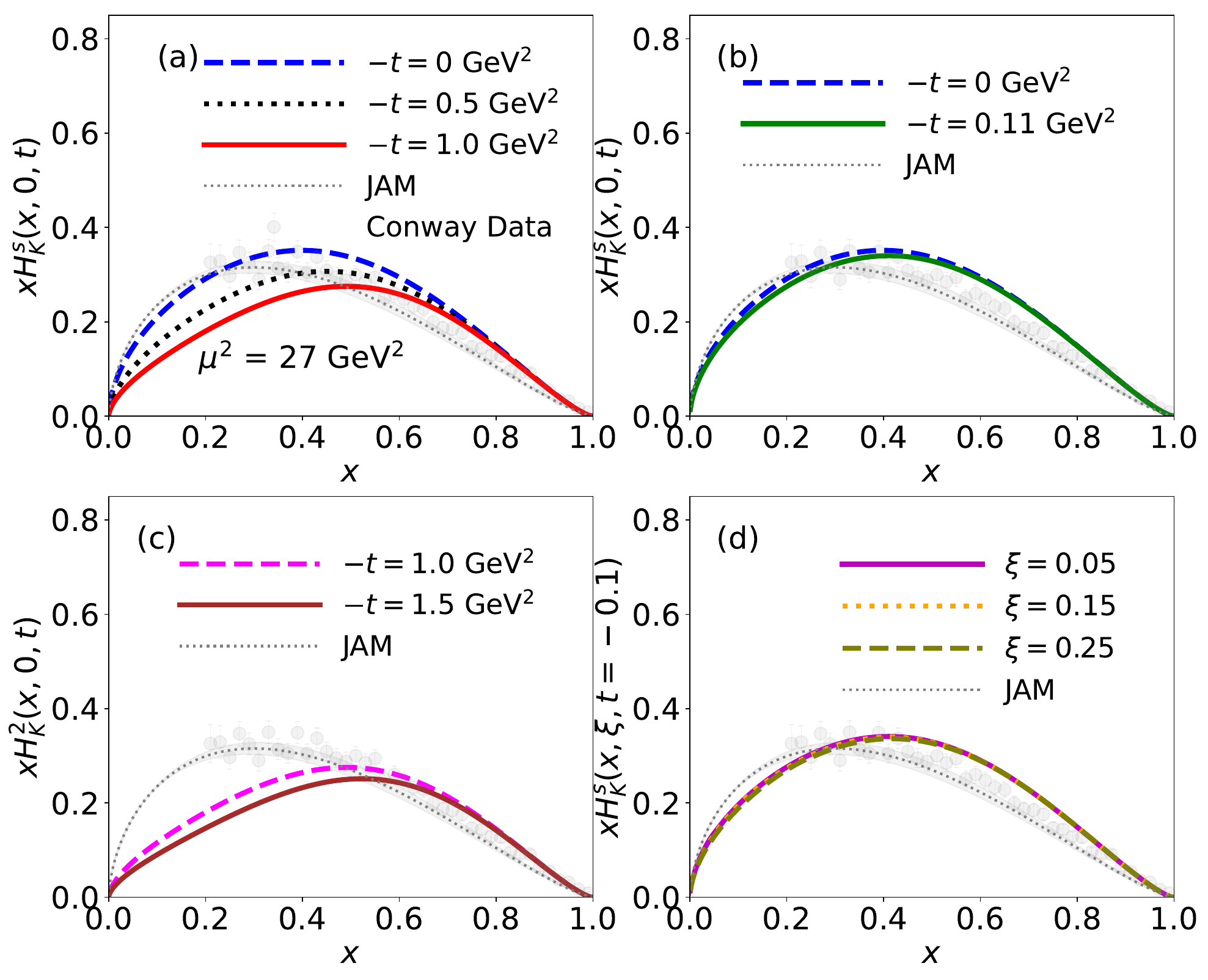} 
\caption{\label{fig5a} Same as in Fig.~\ref{fig5}, but for the strange quark.} 
\end{figure}

In addition to the up-valence-quark distribution, we present our results for the kaon strange-valence-quark distribution for various values of $-t$ and $\xi$ at the scale $\mu^2=27~\mathrm{GeV}^2$, as shown in Figs.~\ref{fig5a}(a)-\ref{fig5a}(d). Figures~\ref{fig5a}(a) and \ref{fig5a}(b) show the strange-valence-quark distributions for $-t=0$, $0.11$, $0.5$, and $1.0~\mathrm{GeV}^2$. We find that the distribution is progressively suppressed with increasing $-t$, with the differences becoming more pronounced in the region $0\leq x\leq0.7$. For $x>0.7$, the dependence on $-t$ is much weaker, and the differences among the kaon strange-valence-quark distributions become negligible.

In Fig.~\ref{fig5a}(c), we consider the larger momentum transfer, $-t=1.5~\mathrm{GeV}^2$. We find that the kaon strange-valence-quark distribution is further suppressed with increasing $-t$ in the region $0\leq x\leq0.7$, whereas for $x>0.7$, the dependence on $-t$ is weak and the differences among the strange-valence-quark distributions become negligible.

The kaon strange-valence-quark distributions for $\xi=0.05$, $0.15$, and $0.25$ at the scale $\mu^2=27~\mathrm{GeV}^2$ are shown in Fig.~\ref{fig5a}(d). We find that the strange-valence-quark distribution exhibits only a weak dependence on $\xi$, decreasing slightly as $\xi$ increases. In contrast to the up-valence-quark distribution shown in Fig.~\ref{fig5}(d), no crossing point is observed near $x\simeq0.2$. Instead, the strange-valence-quark distribution decreases gradually with increasing $\xi$ in the region $x<0.7$, while the differences among the distributions become negligible for $x>0.7$.
\begin{figure}[t]
\centering
\includegraphics[width=1\columnwidth]{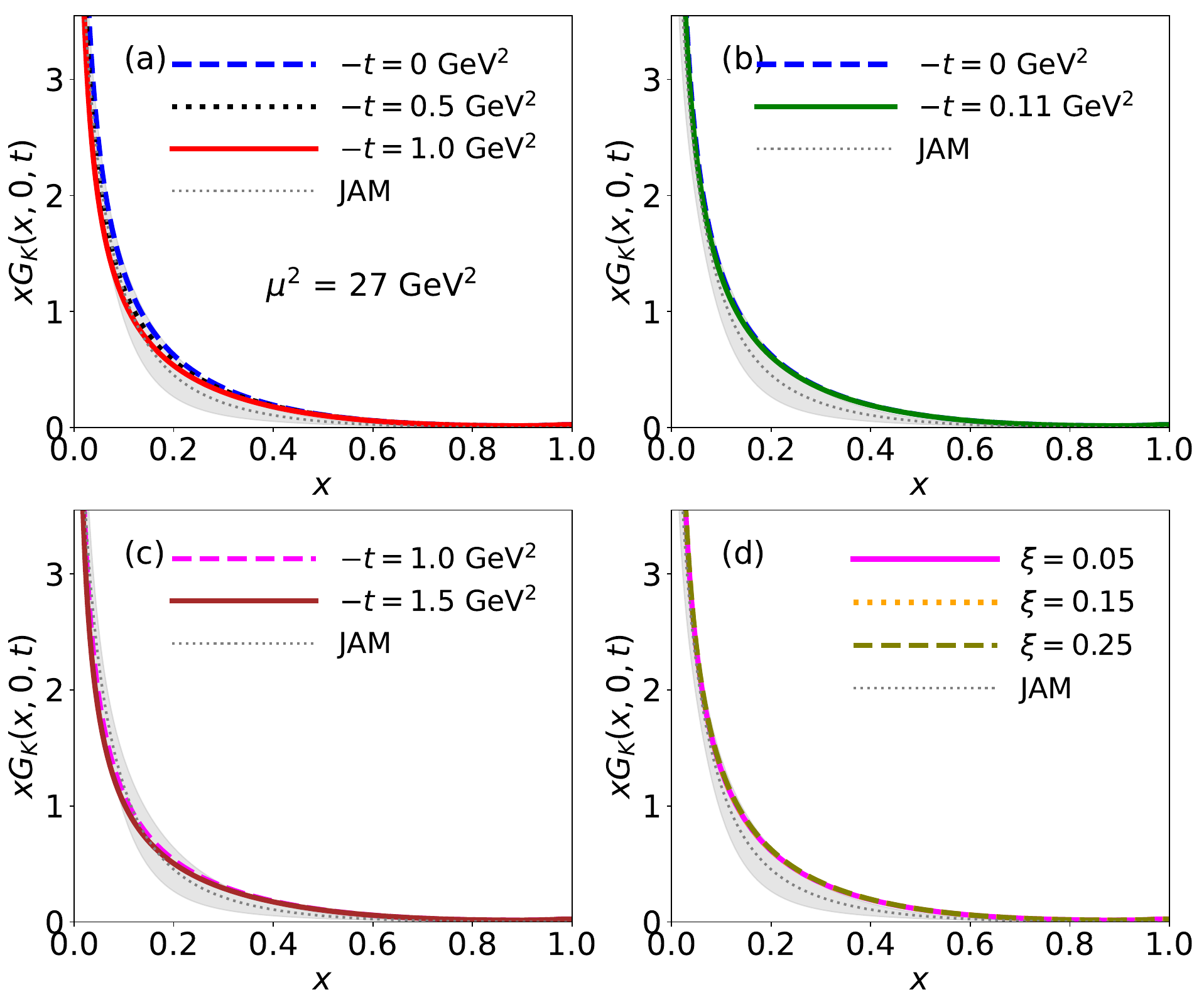} 
\caption{\label{fig6} Gluon distribution of the kaon at the scale $\mu^2 =$ 27 GeV$^2$. For reference, we also show the pion gluon distribution from the JAM QCD analysis at the same scale~\cite{Barry:2021osv}.} 
\end{figure}

In Figs.~\ref{fig6}(a)-\ref{fig6}(d), we present the kaon gluon distributions as functions of $x$ for different values of $-t$ and $\xi$. Figure~\ref{fig6}(a) shows that the kaon gluon distribution decreases with increasing $-t$ in the region $0\leq x\leq0.4$, whereas for $x>0.4$, the dependence on $-t$ is weak and the differences among the distributions become negligible. A similar behavior is observed in Fig.~\ref{fig6}(b). The suppression becomes more pronounced at $-t=1.5~\mathrm{GeV}^2$, as shown in Fig.~\ref{fig6}(c). For comparison, we also include the pion gluon distributions in Figs.~\ref{fig6}(a)-\ref{fig6}(d), given the lack of experimental data and lattice-QCD results for the kaon gluon distribution.
\begin{figure}[t]
\centering
\includegraphics[width=1\columnwidth]{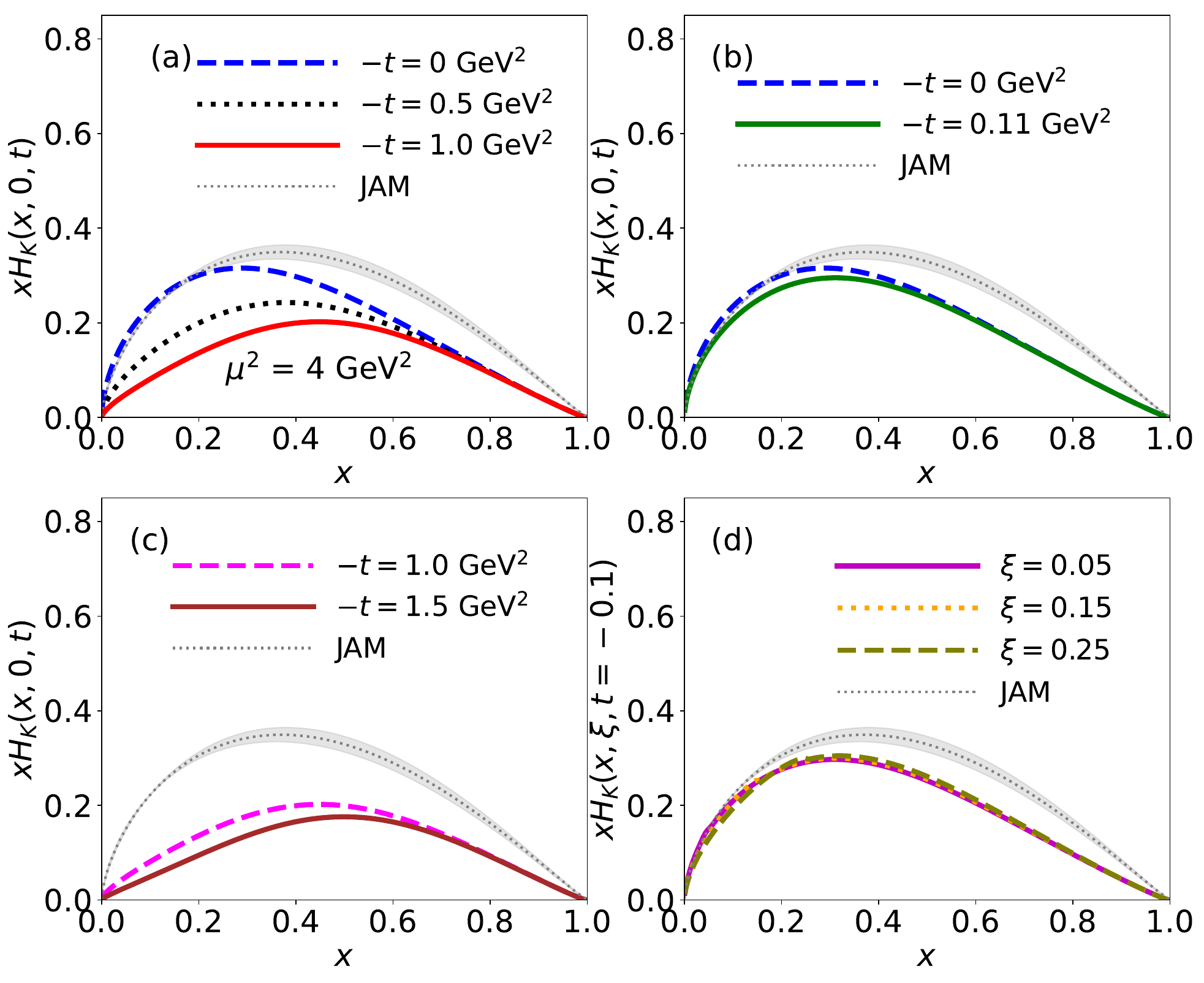} 
\caption{\label{fig7} Same as in Fig.~\ref{fig5}, but for the scale $\mu^2 = 4$ GeV$^2$ in comparison with the JAM QCD analysis for the pion up valence-quark distribution at the same scale~\cite{Barry:2021osv}. } 
\end{figure}

We also compute the kaon up-valence-quark distributions at fixed $\xi=0$ for $-t=0$, $0.11$, $0.5$, $1.0$, and $1.5~\mathrm{GeV}^2$, and at fixed $-t=0.1~\mathrm{GeV}^2$ for $\xi=0.05$, $0.15$, and $0.25$, at the scale $\mu^2=4~\mathrm{GeV}^2$, as shown in Figs.~\ref{fig7}(a)-\ref{fig7}(d). We find that the kaon up-valence-quark distribution is gradually suppressed with increasing $-t$ in the region $0\leq x\leq0.7$, whereas for $x>0.7$, the dependence on $-t$ is weak and the suppression is negligible. A similar behavior is observed in Fig.~\ref{fig7}(b), with the suppression becoming more pronounced at $-t=1.5~\mathrm{GeV}^2$.

Figure~\ref{fig7}(d) shows that the kaon up-valence-quark distribution decreases with increasing $\xi$ for $0\leq x\leq0.2$, whereas it increases with increasing $\xi$ for $x>0.2$.
\begin{figure}[t]
\centering
\includegraphics[width=1\columnwidth]{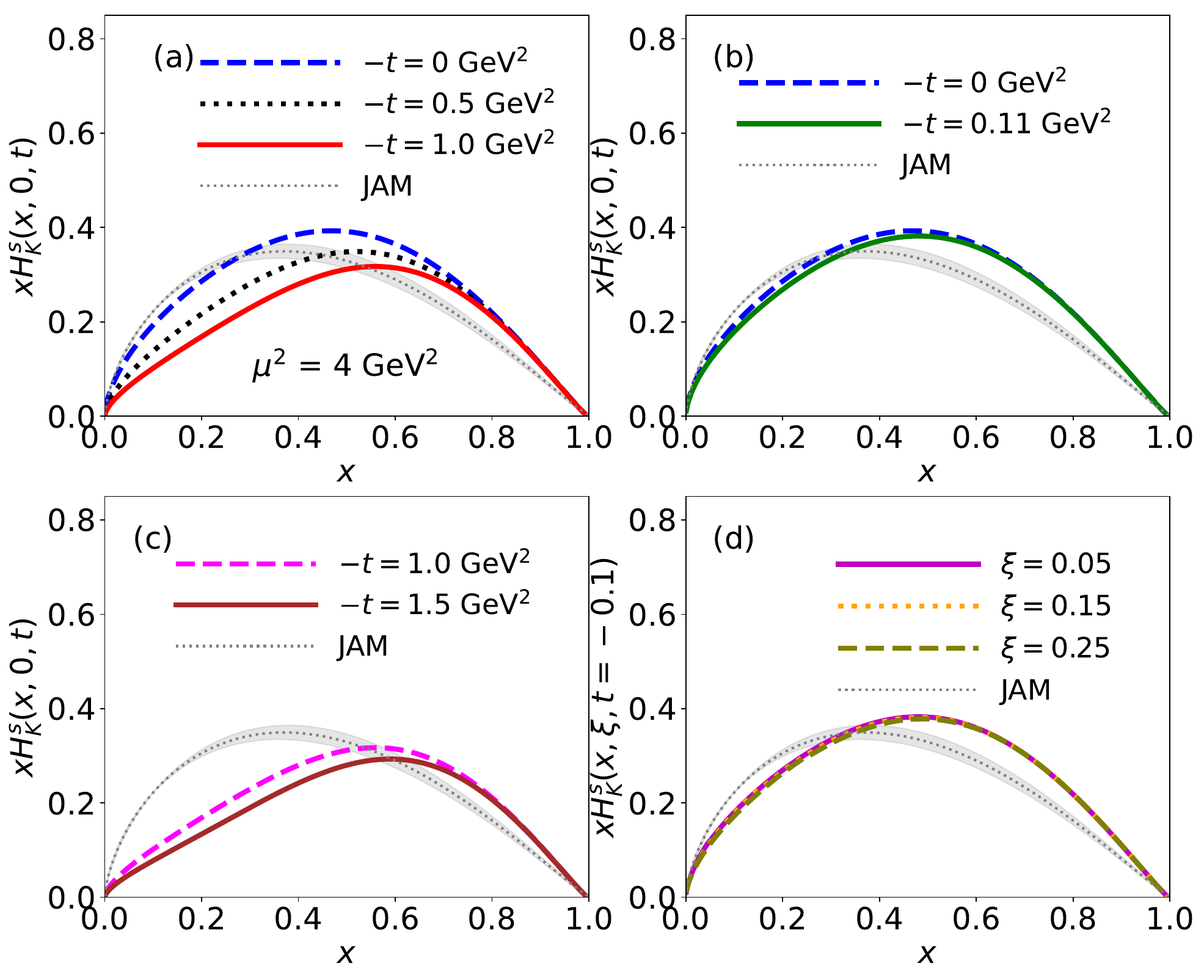} 
\caption{\label{fig7a} Same as in Fig.~\ref{fig6}, but for the strange quark. } 
\end{figure}

The kaon strange-valence-quark distributions at fixed $\xi=0$ for different values of $-t$ at the scale $\mu^2=4~\mathrm{GeV}^2$ are shown in Fig.~\ref{fig7a}(a). We find that the strange-valence-quark distribution decreases with increasing $-t$ in the region $0\leq x\leq0.7$, whereas the dependence on $-t$ becomes negligible for $x>0.7$. A similar behavior is observed in Fig.~\ref{fig7a}(b), where the distribution decreases with increasing $-t$ for $0\leq x\leq0.4$, while the differences among the distributions become negligible for $x>0.4$. The suppression becomes more pronounced at larger values of $-t$, as shown in Fig.~\ref{fig7a}(c).

In Fig.~\ref{fig7a}(d), we present the kaon strange-valence-quark distribution at fixed $-t=0.1~\mathrm{GeV}^2$ for $\xi=0.05$, $0.15$, and $0.25$ at the scale $\mu^2=4~\mathrm{GeV}^2$. We find that the strange-valence-quark distribution exhibits a slight decrease with increasing $\xi$. Similar to the results at $\mu^2=27~\mathrm{GeV}^2$, we find no evidence of a crossing point near $x\simeq0.2$ among the strange-valence-quark distributions for different values of $\xi$ at $\mu^2=4~\mathrm{GeV}^2$.
\begin{figure}[t]
\centering
\includegraphics[width=1\columnwidth]{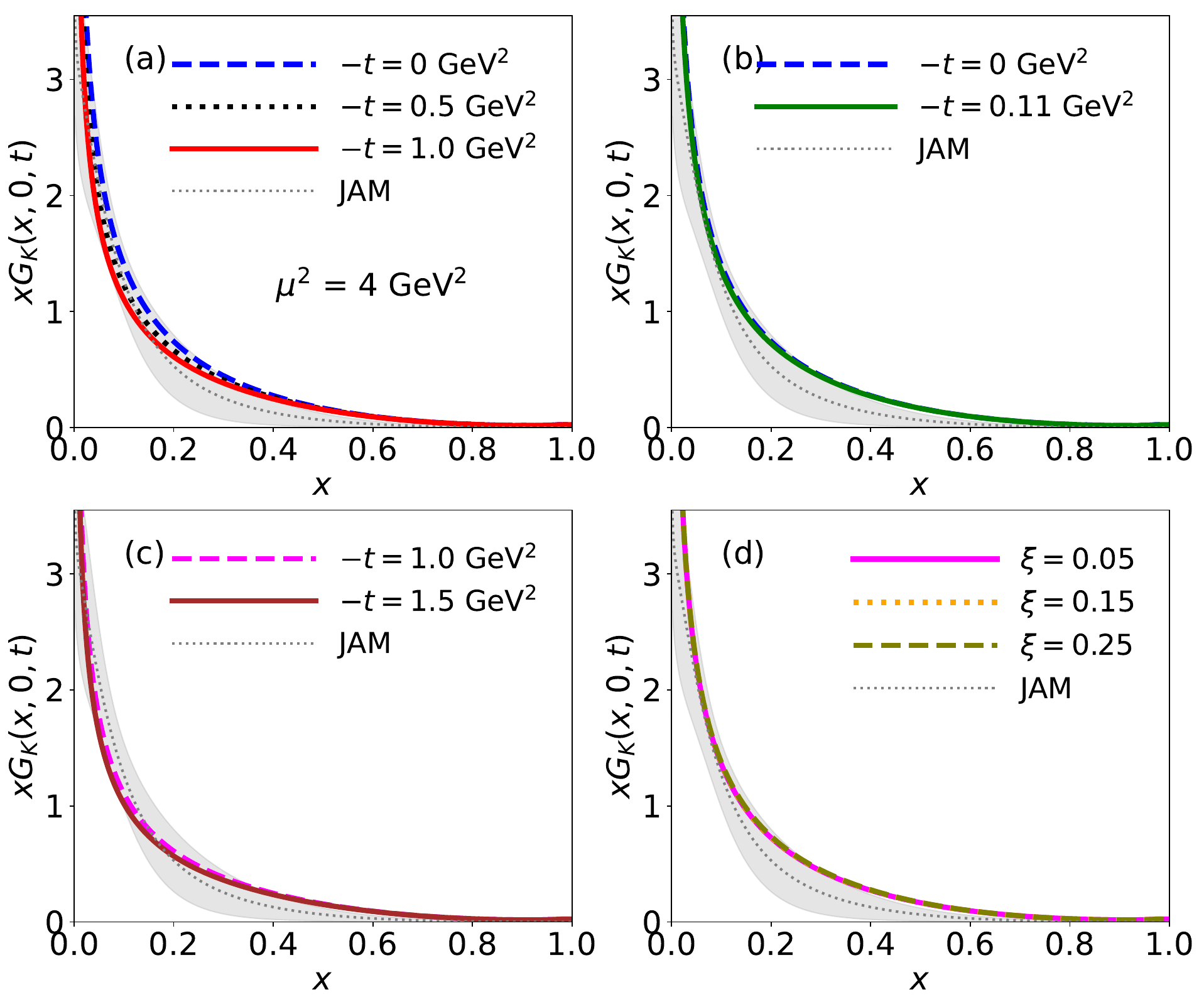} 
\caption{\label{fig8} Same as in Fig.~\ref{fig7}, but for the renormalization scale $\mu^2 = 4$ GeV$^2$ in comparison with the JAM QCD analysis with the same scale~\cite{Barry:2021osv}.} 
\end{figure}

The kaon gluon distributions at fixed $\xi=0$ for $-t=0$, $0.11$, $0.5$, $1.0$, and $1.5~\mathrm{GeV}^2$ at the scale $\mu^2=4~\mathrm{GeV}^2$ are shown in Figs.~\ref{fig8}(a)-\ref{fig8}(c). We find that the kaon gluon distribution decreases with increasing $-t$. This suppression becomes more pronounced at larger values of $-t$, as shown in Fig.~\ref{fig8}(c).

In Fig.~\ref{fig8}(d), we present our results for the kaon gluon distributions at $\xi=0.05$, $0.15$, and $0.25$. We find that the gluon distributions exhibit only a weak dependence on $\xi$ at the scale $\mu^2=4~\mathrm{GeV}^2$.

\subsection{Results for the pion GFFs}
\begin{figure}[t]
\centering
\includegraphics[width=1\columnwidth]{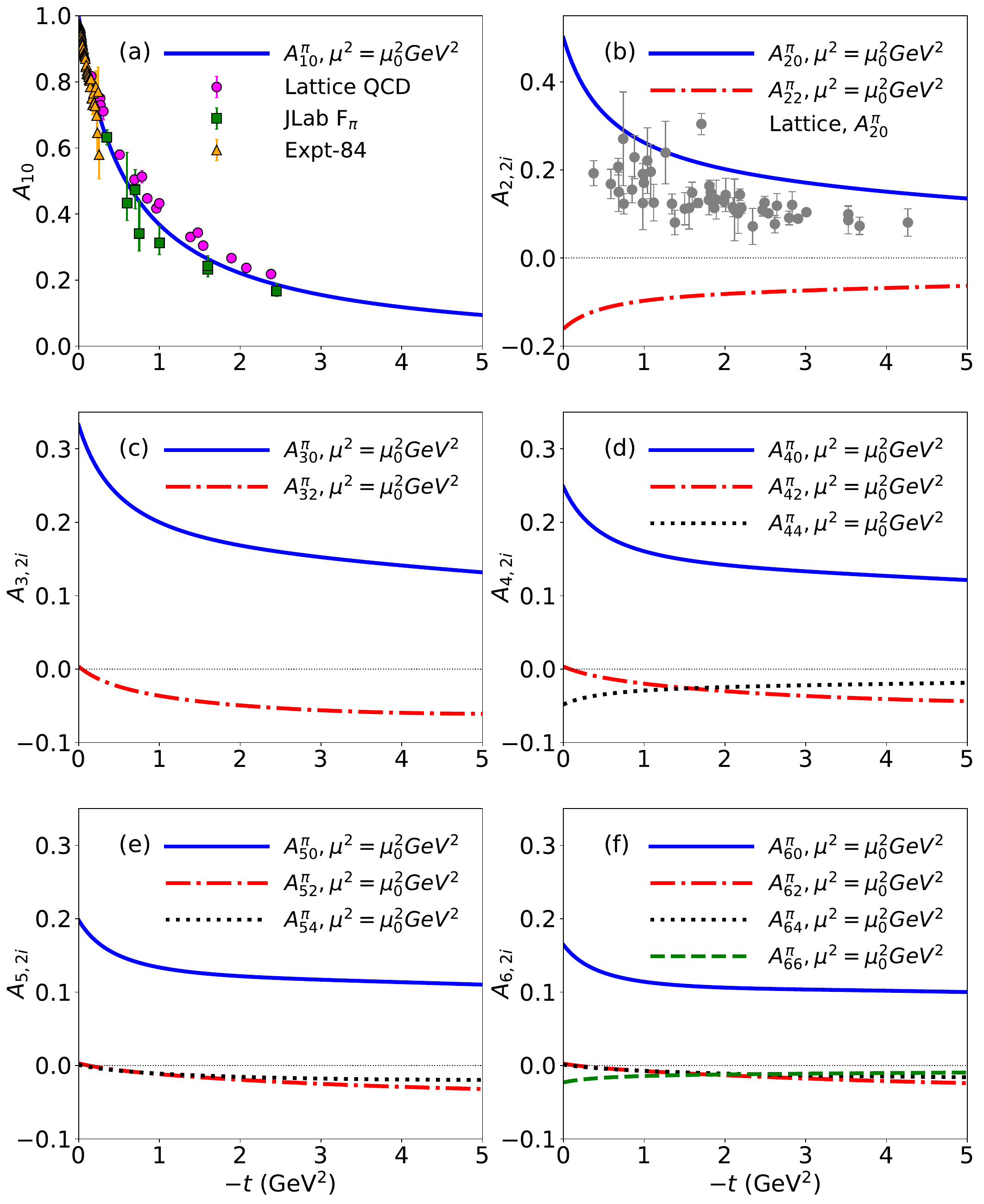} 
\caption{\label{fig9} Pion generalized form factors (GFFs) with  (a) $A^{\pi}_{10}(t)$, (b) $A^{\pi}_{20}(t)$ and $A^{\pi}_{22}(t)$, (c) $A^{\pi}_{30}(t)$ and $A^{\pi}_{32}(t)$, (d) $A^{\pi}_{40}(t)$, $A^{\pi}_{42}(t)$, and $A^{\pi}_{44}(t)$, (e) $A^{\pi}_{50}(t)$, $A^{\pi}_{52}(t)$, and $A^{\pi}_{54}(t)$, and (f) $A^{\pi}_{60}(t)$, $A^{\pi}_{62}(t)$, $A^{\pi}_{64}(t)$, and $A^{\pi}_{66}(t)$. Note that the GFFs are evaluated at the initial scale $\mu^2=\mu_0^2$. The lattice-QCD results for $A_{20}^{\pi}(t)$ at $\mu^2 =$ 4 GeV$^2$ from Refs.~\cite{Brommel:2007zz,Broniowski:2008hx} are shown for reference. The lattice QCD results for $A_{10}^{\pi}$ at $\mu^2 =$ 4 GeV$^2$ are taken from Ref.~\cite{Alexandrou:2021ztx}, while the experimental data are taken from Refs.~\cite{Amendolia:1984nz,NA7:1986vav,JeffersonLabFpi-2:2006ysh,JeffersonLabFpi:2007vir,JeffersonLab:2008jve,JeffersonLab:2008gyl}.} 
\end{figure}

In Fig.~\ref{fig9}(a), we present our results for the pion vector form factor, $A_{10}^{\pi}(t)=F_{\pi}^{V}(t)$, together with lattice-QCD results~\cite{Alexandrou:2021ztx} and experimental data~\cite{E615:1989bda}. Our results are in good agreement with both the lattice-QCD calculation and the experimental data over the displayed range of momentum transfer. The form factor satisfies the normalization condition $ A_{10}^{\pi}(0)=1$, as required by charge conservation. We also find that $A_{10}^{\pi}(t)$ decreases monotonically with increasing $-t$. In the asymptotic region, the form factor exhibits the expected perturbative-QCD counting-rule behavior, $A_{10}^{\pi}(t)\sim 1/Q^2$, where $Q^2=-t$.

Figure~\ref{fig9}(b) shows our results for the pion generalized form factors $A_{20}^{\pi}(t)$ and $A_{22}^{\pi}(t)$ as functions of the momentum transfer $-t$. We find that $A_{20}^{\pi}(t)$ is positive and decreases with increasing $-t$, in agreement with the lattice results of Ref.~\cite{Delmar:2024vxn} and theoretical prediction results in Refs.~\cite{Broniowski:2008hx,Fanelli:2016aqc}. In contrast, $A_{22}^{\pi}(t)$ is negative and increases with increasing $-t$, consistent with the results reported in Ref.~\cite{Fanelli:2016aqc}.

Figure~\ref{fig9}(c) shows the form factors $A_{30}^{\pi}(t)$ and $A_{32}^{\pi}(t)$ as functions of the momentum transfer $-t$. We find that $A_{30}^{\pi}(t)$ is positive and decreases with increasing $-t$, which is in good agreement with the prediction results in Ref.~\cite{Broniowski:2008hx}. In contrast, $A_{32}^{\pi}(t)$ is negative for finite $-t$, satisfies the normalization condition $A_{32}^{\pi}(0)=0$, and decreases further with increasing $-t$. Our results on $A_{32}^{\pi}(t)$ are also consistent with that obtained in Ref.~\cite{Broniowski:2008hx}.

Figure~\ref{fig9}(d) shows our results for the form factors $A_{40}^{\pi}(t)$, $A_{42}^{\pi}(t)$, and $A_{44}^{\pi}(t)$ as functions of the momentum transfer $-t$. We find that $A_{40}^{\pi}(t)$ is positive and decreases with increasing $-t$, which is consistent with that in Ref.~\cite{Broniowski:2008hx}. In contrast, both $A_{42}^{\pi}(t)$ and $A_{44}^{\pi}(t)$ are negative. The magnitude of $A_{42}^{\pi}(t)$ increases with increasing $-t$, whereas $A_{44}^{\pi}(t)$ becomes less negative as $-t$ increases. These findings are consistent with those obtained in Ref.~\cite{Broniowski:2008hx}.

Furthermore, Fig.~\ref{fig9}(e) shows our results for the higher order form factors $A_{50}^{\pi}(t)$, $A_{52}^{\pi}(t)$, and $A_{54}^{\pi}(t)$ as functions of the momentum transfer $-t$. We find that $A_{50}^{\pi}(t)$ is positive and decreases with increasing $-t$. Both $A_{52}^{\pi}(t)$ and $A_{54}^{\pi}(t)$ are negative and decrease with increasing $-t$.

The results for the form factors $A_{60}^{\pi}(t)$, $A_{62}^{\pi}(t)$, $A_{64}^{\pi}(t)$, and $A_{66}^{\pi}(t)$ as functions of $-t$ are shown in Fig.~\ref{fig9}(f). We find that $A_{60}^{\pi}(t)$ is positive and decreases with increasing $-t$. In contrast, $A_{62}^{\pi}(t)$, $A_{64}^{\pi}(t)$, and $A_{66}^{\pi}(t)$ are negative. Both $A_{62}^{\pi}(t)$ and $A_{64}^{\pi}(t)$ decrease with increasing $-t$, whereas $A_{66}^{\pi}(t)$ increases with increasing $-t$.

\subsection{Results for the kaon GFFs}
\begin{figure}[ht]
\centering
\includegraphics[width=1\columnwidth]{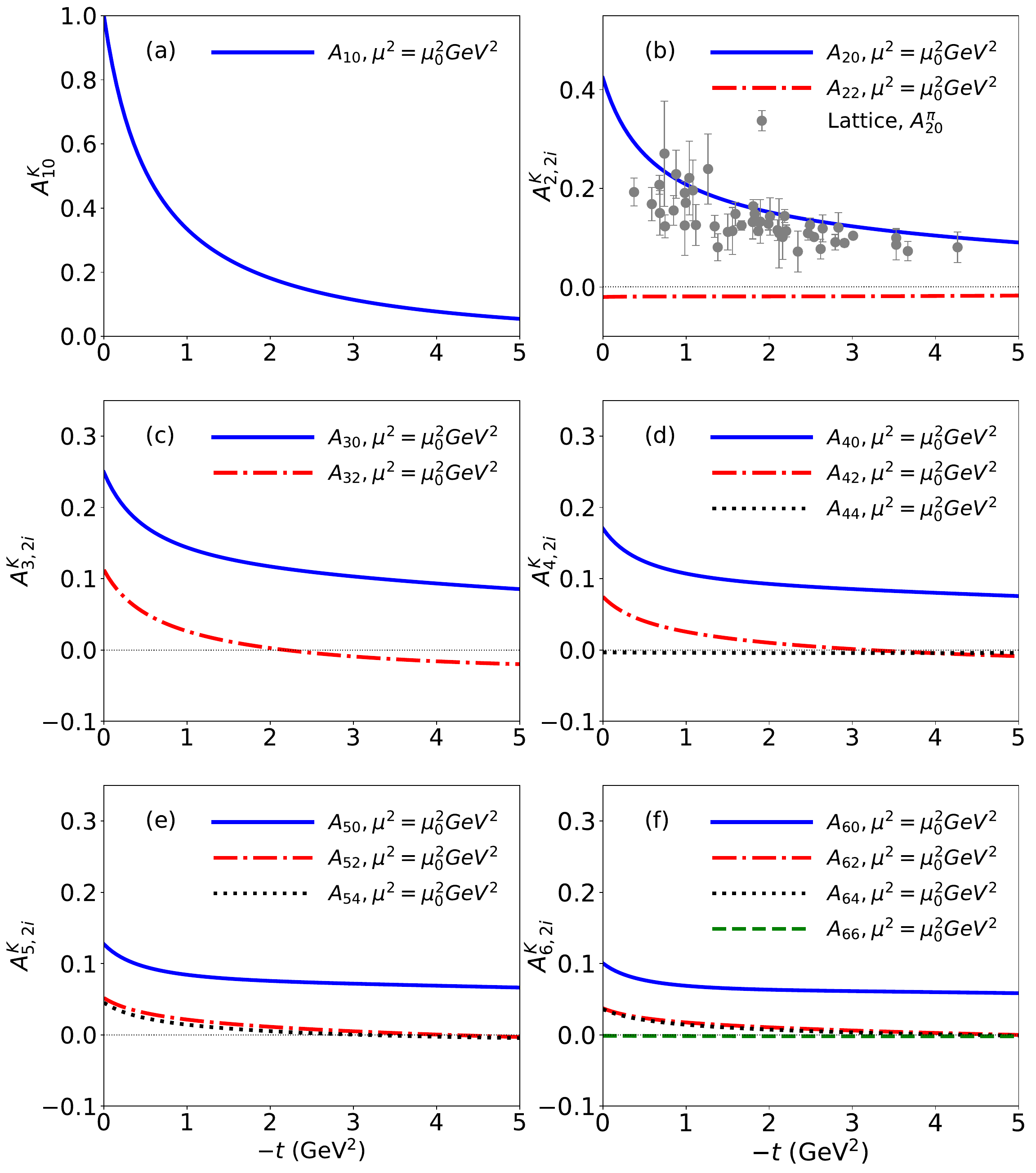} 
\caption{\label{fig10} Kaon generalized form factors (GFFs) for the up quark with (a) $A^{uK}_{10}(t)$, (b) $A^{uK}_{20}(t)$ and $A^{uK}_{22}(t)$, (c) $A^{uK}_{30}(t)$ and $A^{uK}_{32}(t)$, (d) $A^{uK}_{40}(t)$, $A^{uK}_{42}(t)$, and $A^{uK}_{44}(t)$, (e) $A^{uK}_{50}(t)$, $A^{uK}_{52}(t)$, and $A^{uK}_{54}(t)$, and (f) $A^{uK}_{60}(t)$, $A^{uK}_{62}(t)$, $A^{uK}_{64}(t)$, and $A^{uK}_{66}(t)$. Note that the GFFs are computed with the initial scale $\mu^2 = \mu_0^2$. The lattice-QCD results for $A_{20}^{\pi}(t)$ at the scale $\mu^2 =$ 4 GeV$^2$ from Refs.~\cite{Brommel:2007zz,Broniowski:2008hx} are shown for reference only.} 
\end{figure}

Next, we present the results for the kaon higher-order generalized form factors in Figs.~\ref{fig10}(a)-\ref{fig10}(f). In Fig.~\ref{fig10}(a), we show the form factor $A_{10}^{uK}(t)$ as a function of the momentum transfer $-t$. We find that $A_{10}^{uK}(t)$ decreases with increasing $-t$. At $t=0$, the normalization condition $A_{10}^{uK}(0)=1$ is satisfied, as required by charge conservation. Our result is consistent with that of Ref.~\cite{Son:2024uet}.

Figure~\ref{fig10}(b) shows the results for $A_{20}^{uK}(t)$ and $A_{22}^{uK}(t)$, together with the lattice-QCD results for $A_{20}(t)$~\cite{Brommel:2007zz}. We find that $A_{20}^{uK}(t)$ decreases with increasing $-t$, whereas $A_{22}^{uK}(t)$ is negative and exhibits a nearly constant behavior over the range of $-t$ considered. At the scale $\mu^2=4~\mathrm{GeV}^2$, our results for $A_{20}^{uK}(t)$ are in good agreement with the corresponding lattice-QCD results~\cite{Brommel:2007zz}.

The results for the form factors $A_{30}^{uK}(t)$ and $A_{32}^{uK}(t)$ as functions of $-t$ are shown in Fig.~\ref{fig10}(c). Both form factors decrease with increasing $-t$; however, for $-t\gtrsim 2~\mathrm{GeV}^2$, $A_{32}^{uK}(t)$ becomes negative. Similar behavior is observed for $A_{40}^{uK}(t)$ and $A_{42}^{uK}(t)$, as shown in Fig.~\ref{fig10}(d), both of which decrease with increasing $-t$. In contrast, $A_{44}^{uK}(t)$ is negative and exhibits a nearly constant behavior over the range of $-t$ considered. 

In Fig.~\ref{fig10}(e), we show our results for the form factors $A_{50}^{uK}(t)$, $A_{52}^{uK}(t)$, and $A_{54}^{uK}(t)$ as functions of $-t$. We find that all three form factors decrease with increasing $-t$. In particular, $A_{52}^{uK}(t)$ and $A_{54}^{uK}(t)$ approach zero at $-t\simeq 2.5~\mathrm{GeV}^2$.

Figure~\ref{fig10}(f) shows the form factors $A_{60}^{uK}(t)$, $A_{62}^{uK}(t)$, $A_{64}^{uK}(t)$, and $A_{66}^{uK}(t)$. We find that $A_{60}^{uK}(t)$, $A_{62}^{uK}(t)$, and $A_{64}^{uK}(t)$ decrease with increasing $-t$, whereas $A_{66}^{uK}(t)$ exhibits a monotonic behavior and approaches zero over the range of $-t$ considered.
\begin{figure}[t]
\centering
\includegraphics[width=1\columnwidth]{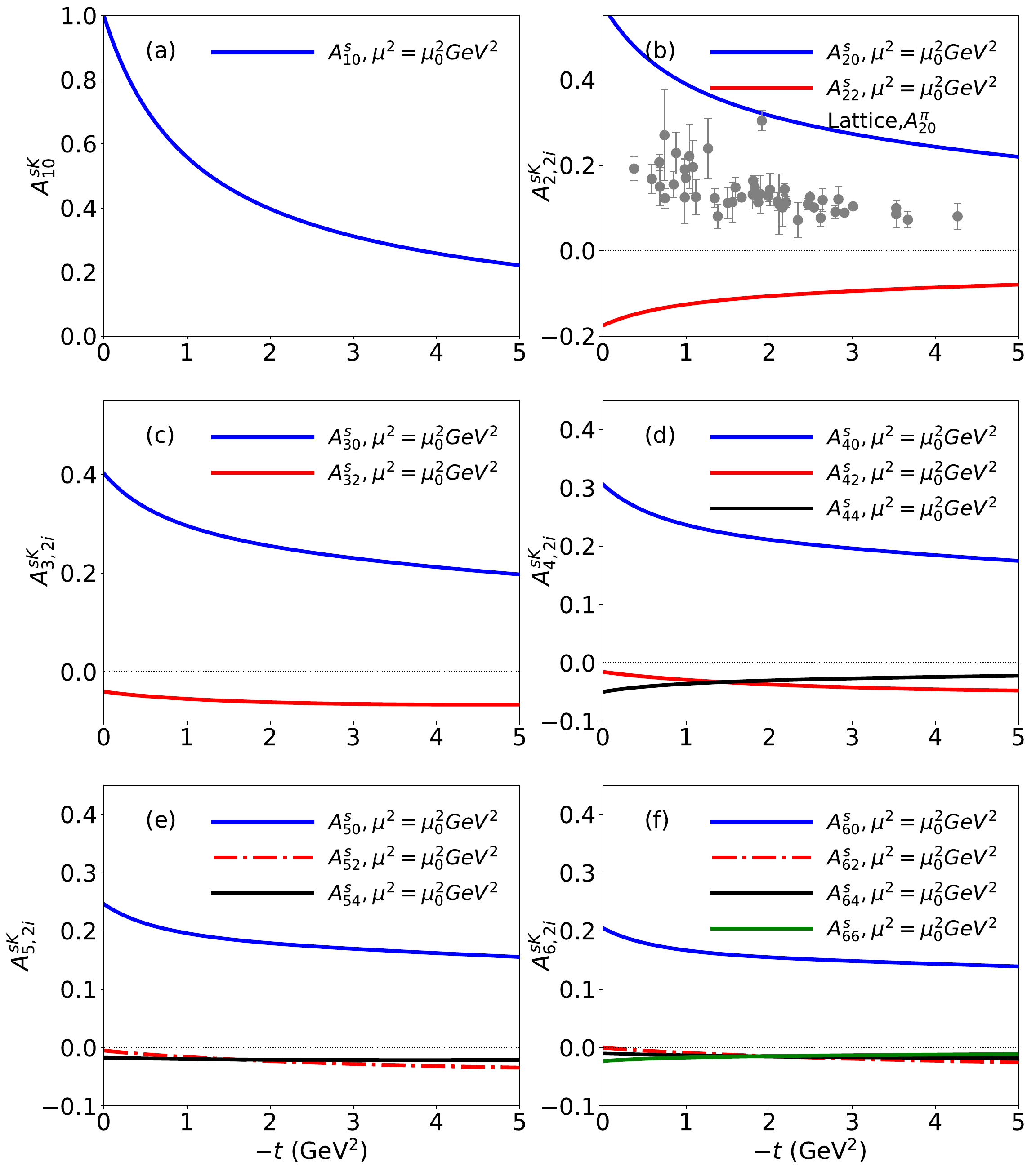} 
\caption{\label{fig11} Kaon generalized form factors (GFFs) for the strange quark with (a) $A^{sK}_{10}(t)$, (b) $A^{sK}_{20}(t)$ and $A^{sK}_{22}(t)$, (c) $A^{sK}_{30}(t)$ and $A^{sK}_{32}(t)$, (d) $A^{sK}_{40}(t)$, $A^{sK}_{42}(t)$, and $A^{sK}_{44}(t)$, (e) $A^{sK}_{50}(t)$, $A^{sK}_{52}(t)$, and $A^{sK}_{54}(t)$, and (f) $A^{sK}_{60}(t)$, $A^{sK}_{62}(t)$, $A^{sK}_{64}(t)$, and $A^{sK}_{66}(t)$. Note that the kaon strange GFFs are computed with the initial scale $\mu^2 = \mu_0^2$. The lattice-QCD results for $A_{20}^{\pi}(t)$ at the scale $\mu^2=$ 4 GeV$^2$ from Refs.~\cite{Brommel:2007zz,Broniowski:2008hx} are shown for reference only.} 
\end{figure}

In addition to the generalized form factors for the up quark, we present the corresponding results for the strange quark in the kaon in Figs.~\ref{fig11}(a)-\ref{fig11}(f). Figure~\ref{fig11}(a) shows the strange-quark form factor $A_{10}^{sK}(t)=F_K^s(Q^2)$, with $t=-Q^2$, which decreases with increasing $-t$. At $t=0$, the normalization condition $A_{10}^{sK}(0)=1$ is satisfied, as required by charge conservation. We note that this form factor is evaluated numerically at the initial scale $\mu^2=\mu_0^2$.

The results for the form factors $A_{20}^{sK}(t)$ and $A_{22}^{sK}(t)$ as functions of $-t$ are shown in Fig.~\ref{fig11}(b). We find that $A_{20}^{sK}(t)$ decreases with increasing $-t$. In contrast, $A_{22}^{sK}(t)$ is negative and increases with increasing $-t$.

Similar behavior is observed for $A_{30}^{sK}(t)$ and $A_{32}^{sK}(t)$, as shown in Fig.~\ref{fig11}(c). Both form factors decrease with increasing $-t$, while $A_{32}^{sK}(t)$ remains negative over the range considered.

In Fig.~\ref{fig11}(d), we show our results for the form factors $A_{40}^{sK}(t)$, $A_{42}^{sK}(t)$, and $A_{44}^{sK}(t)$ as functions of $-t$. We find that both $A_{40}^{sK}(t)$ and $A_{42}^{sK}(t)$ decrease with increasing $-t$. The form factor $A_{42}^{sK}(t)$ is negative, as is $A_{44}^{sK}(t)$. However, these form factors exhibit distinct $-t$ dependencies, with $A_{44}^{sK}(t)$ increasing as $-t$ increases.

We further present our results for the form factors $A_{50}^{sK}(t)$, $A_{52}^{sK}(t)$, and $A_{54}^{sK}(t)$ as functions of $-t$ in Fig.~\ref{fig11}(d). We find that both $A_{52}^{sK}(t)$ and $A_{54}^{sK}(t)$ are negative and exhibit distinct $-t$ dependencies. In particular, $A_{52}^{sK}(t)$ decreases with increasing $-t$, whereas $A_{54}^{sK}(t)$ remains nearly constant over the range considered. Similar to $A_{52}^{sK}(t)$, $A_{50}^{sK}(t)$ also decreases with increasing $-t$.

The results for the form factors $A_{60}^{sK}(t)$, $A_{62}^{sK}(t)$, $A_{64}^{sK}(t)$, and $A_{66}^{sK}(t)$ are shown in Fig.~\ref{fig11}(f). We find that both $A_{60}^{sK}(t)$ and $A_{62}^{sK}(t)$ decrease with increasing $-t$. However, they differ in sign: $A_{60}^{sK}(t)$ remains positive, whereas $A_{62}^{sK}(t)$ is negative. Similar to $A_{62}^{sK}(t)$, the form factors $A_{64}^{sK}(t)$ and $A_{66}^{sK}(t)$ are also negative. Their $-t$ dependence, however, differs, with $A_{64}^{sK}(t)$ exhibiting a relatively weak dependence on $-t$.

\subsection{Results for the evolved pion and kaon GFFs}
\begin{figure*}[t]
\centering
\includegraphics[width=1\columnwidth]{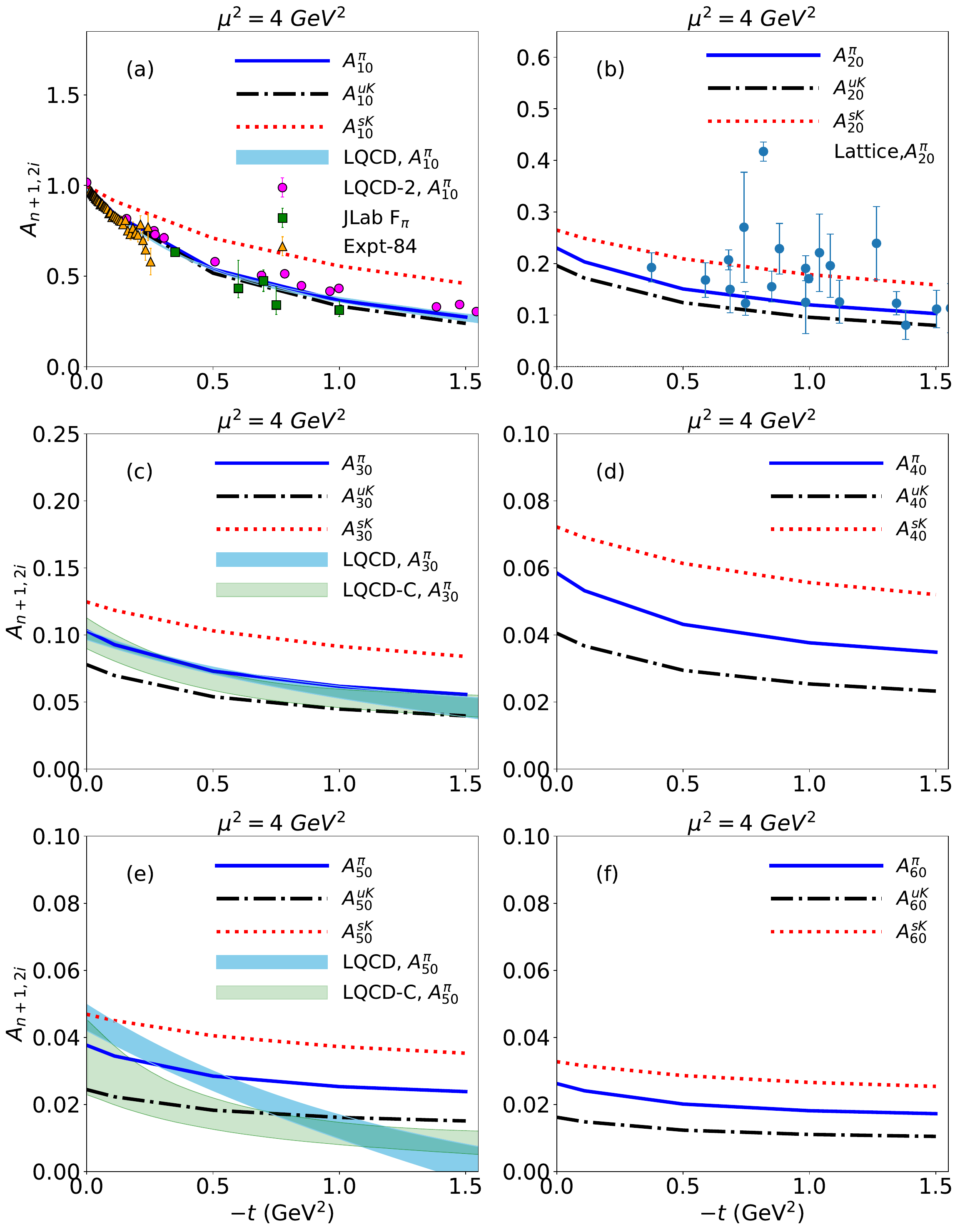} 
\includegraphics[width=1\columnwidth]{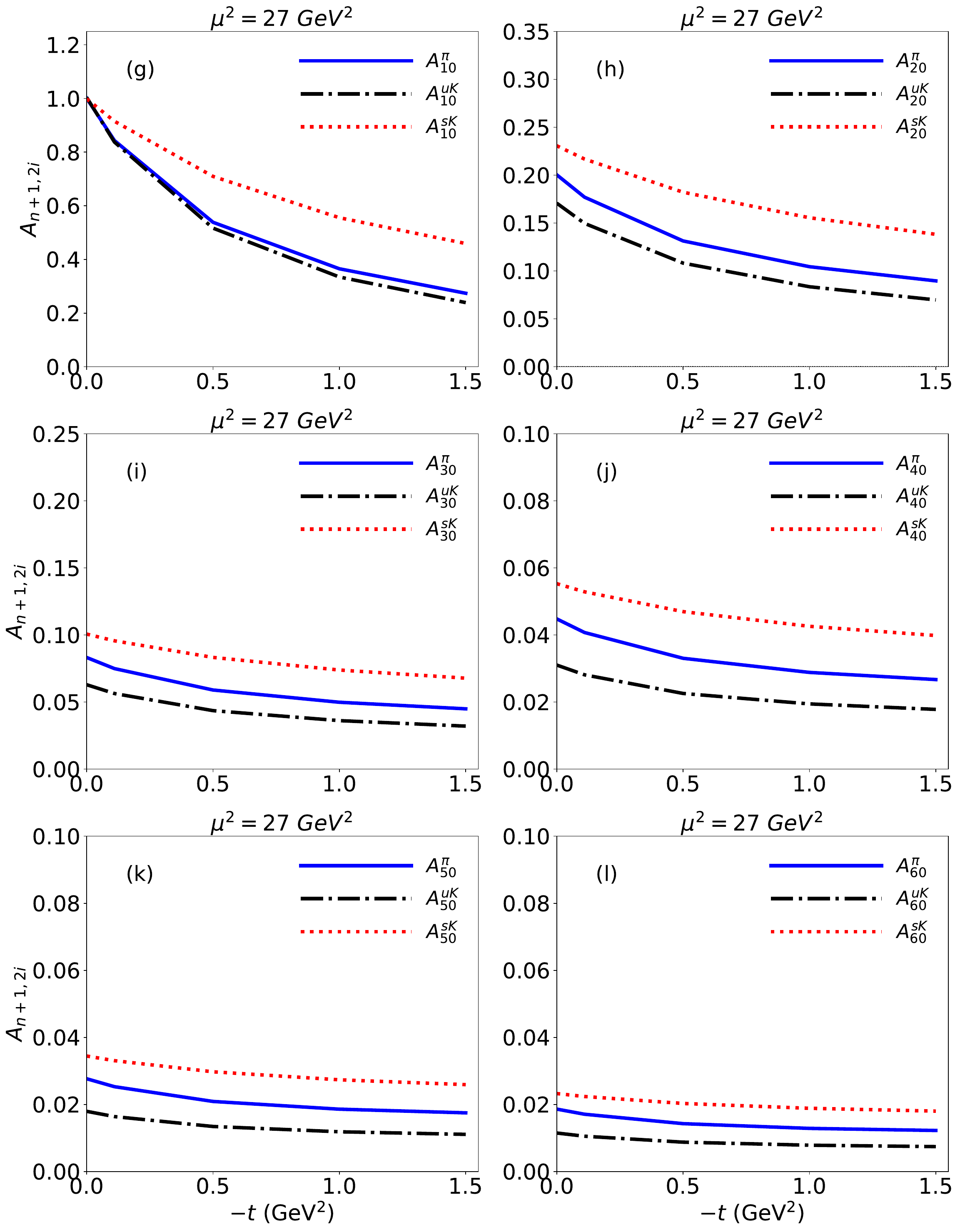}
\caption{\label{fig12} Kaon and pion generalized form factors (GFFs) at the scales $\mu^2 =$ 4 GeV$^2$ (left panel) and 27 GeV$^2$ (right panel). The lattice-QCD results for $A_{20}^{\pi}(t)$ at $\mu^2 =$ 4 GeV$^2$ from Refs.~\cite{Brommel:2007zz,Broniowski:2008hx} are shown for comparison only. The lattice QCD results for $A_{10}^{\pi}$(t) at $\mu^2 =$ 4 GeV$^2$ are taken from Refs.~\cite{Alexandrou:2021ztx,Gao:2025inf}, while the experimental data are taken from Refs.~\cite{Amendolia:1984nz,NA7:1986vav,JeffersonLabFpi-2:2006ysh,JeffersonLabFpi:2007vir,JeffersonLab:2008jve,JeffersonLab:2008gyl}.} 
\end{figure*}

To facilitate a quantitative comparison of our GFFs for the pion and kaon with lattice-QCD results, we compute the GFFs at the scales $\mu^2=4$ and $27~\mathrm{GeV}^2$ using the evolved pion and kaon GPDs at both zero and nonzero skewness, as shown in Figs.~\ref{fig12}(a)-\ref{fig12}(l).

Figure~\ref{fig12}(a) shows the results for $A_{10}^{u\pi}(t)$, $A_{10}^{uK}(t)$, and $A_{10}^{sK}(t)$ at the scale $\mu^2=4~\mathrm{GeV}^2$ as functions of $-t$. We find that all three form factors decrease with increasing $-t$. Moreover, $A_{10}^{u\pi}(t)$ is stiffer than $A_{10}^{uK}(t)$ but softer than $A_{10}^{sK}(t)$. The form factors satisfy the normalization condition $A_{10}^{u\pi}(0)=A_{10}^{uK}(0)=A_{10}^{sK}(0)=1$, as required by charge conservation. Also, we find that our result for the $A_{10}^{\pi}$(t) at $\mu^2 =$ 4 GeV$^2$ is in excellent agreement with the lattice QCD result at $\mu^2 =$ 4 GeV$^2$~\cite{Alexandrou:2021ztx,Gao:2025inf} and the experimental data~\cite{Amendolia:1984nz,NA7:1986vav,JeffersonLabFpi-2:2006ysh,JeffersonLabFpi:2007vir,JeffersonLab:2008jve,JeffersonLab:2008gyl}.

In addition to the $A_{10}(t)$ form factors for the pion and kaon, we present the higher-moment form factors $A_{20}^{u\pi}(t)$, $A_{20}^{uK}(t)$, and $A_{20}^{sK}(t)$ at the scale $\mu^2=4~\mathrm{GeV}^2$ in Fig.~\ref{fig12}(b). We find that $A_{20}^{u\pi}(t)$ is larger than $A_{20}^{uK}(t)$ over the entire range of $-t$ considered. In contrast, $A_{20}^{u\pi}(t)$ is softer than $A_{20}^{sK}(t)$. In comparison with the lattice-QCD result~\cite{Brommel:2007zz} and theoretical result~\cite{Broniowski:2008hx}, our result is in good agreement with the lattice calculation over the range of $-t$ considered.

In Fig.~\ref{fig12}(c), we show our results for the form factors $A_{30}^{u\pi}(t)$, $A_{30}^{uK}(t)$, and $A_{30}^{sK}(t)$ as functions of $-t$. We find that $A_{30}^{u\pi}(t)$ is larger than $A_{30}^{uK}(t)$ over the range of $-t$. In comparison with $A_{30}^{sK}(t)$, we find that $A_{30}^{u\pi}(t)$ is smaller than $A_{30}^{sK}(t)$. Our results for $A_{30}^{\pi}(t)$ are in excellent agreement with the lattice QCD result in Ref.~\cite{Gao:2025inf}.

For $A_{40}^{u\pi}(t)$, $A_{40}^{uK}(t)$, and $A_{40}^{sK}(t)$, we find that the generalized form factors exhibit the hierarchy
$ A_{40}^{sK}(t)>A_{40}^{u\pi}(t)>A_{40}^{uK}(t)$ over the entire range of $-t$ considered, as shown in Fig.~\ref{fig12}(d).

Furthermore, we present our results for $A_{10}^{u\pi}(t)$, $A_{10}^{uK}(t)$, and $A_{10}^{sK}(t)$ at the scale $\mu^2=27~\mathrm{GeV}^2$ in Fig.~\ref{fig12}(f). Similar to the GFF results for the pion and kaon at $\mu^2=4~\mathrm{GeV}^2$, we find that $A_{10}^{u\pi}(t)$ is stiffer than $A_{10}^{uK}(t)$ but softer than $A_{10}^{sK}(t)$. We note that all GFFs satisfy the normalization condition at $t=0$, namely,
$A_{10}^{u\pi}(0)=A_{10}^{uK}(0)=A_{10}^{sK}(0)=1$.

In Fig.~\ref{fig12}(g), we show the form factors $A_{20}^{u\pi}(t)$, $A_{20}^{uK}(t)$, and $A_{20}^{sK}(t)$ at the scale $\mu^2=27~\mathrm{GeV}^2$ as functions of $-t$. We find that $A_{20}^{u\pi}(t)$ is larger than $A_{20}^{uK}(t)$ over the entire range of $-t$ considered. In contrast, $A_{20}^{u\pi}(t)$ is smaller than $A_{20}^{sK}(t)$ throughout the same range. Thus, among the three form factors, $A_{20}^{sK}(t)$ exhibits the weakest $-t$ dependence, i.e., it is the stiffest form factor.

The results for $A_{30}^{u\pi}(t)$, $A_{30}^{uK}(t)$, and $A_{30}^{sK}(t)$ as functions of $-t$ at the scale $\mu^2=27~\mathrm{GeV}^2$ are shown in Fig.~\ref{fig12}(h). We find the hierarchy $A_{30}^{sK}(t)>A_{30}^{u\pi}(t)>A_{30}^{uK}(t)$, which is qualitatively similar to the corresponding results at $\mu^2=4~\mathrm{GeV}^2$. However, the magnitudes of the form factors differ between the two scales.

Figure~\ref{fig12}(i) shows the results for $A_{40}^{u\pi}(t)$, $A_{40}^{uK}(t)$, and $A_{40}^{sK}(t)$ as functions of the momentum transfer $-t$ at the scale $\mu^2=27~\mathrm{GeV}^2$. We find that $A_{40}^{sK}(t)$ exhibits the slowest falloff as $-t$ increases, indicating that it is the stiffest among the three form factors. Furthermore, $A_{40}^{u\pi}(t)$ is larger than $A_{40}^{uK}(t)$ over the considered range of $-t$.

We also present our results for the higher-order GFFs of the pion, $A_{50}^{\pi}(t)$ and $A_{60}^{\pi}(t)$, and the kaon, $A_{50}^{uK}(t)$, $A_{50}^{sK}(t)$, $A_{60}^{uK}(t)$, and $A_{60}^{sK}(t)$, at the scales $\mu^2=4$ and $27~\mathrm{GeV}^2$, as shown in Figs.~\ref{fig12}(e)-(f) and \ref{fig12}(k)-(l), respectively. In Fig.~\ref{fig12}(e), we compare our result for $A_{50}^{\pi}(t)$ with the corresponding lattice-QCD result at $\mu^2=4~\mathrm{GeV}^2$. We find a sizable difference between our prediction and the lattice-QCD result~\cite{Gao:2025inf}; this is subject to the large uncertainties associated with the lattice-QCD calculation. Our predictions for the higher-order GFFs $A_{60}^{\pi}(t)$, $A_{60}^{uK}(t)$, and $A_{60}^{sK}(t)$ at $\mu^2=4$ and $27~\mathrm{GeV}^2$ are shown in Figs.~\ref{fig12}(f) and \ref{fig12}(l), respectively.

Overall, we conclude that the evolved pion GFFs at $\mu^2=4$ and $27~\mathrm{GeV}^2$ exhibit the hierarchy
\begin{eqnarray}
A_{n0}^{uK}(t) < A_{n0}^{u\pi}(t) < A_{n0}^{sK} (t), ~~~~~~n =1,2,3,4.
\end{eqnarray}
Thus, $A_{10}^{u\pi}(t)$, $A_{20}^{u\pi}(t)$, $A_{30}^{u\pi}(t)$, and $A_{40}^{u\pi}(t)$ are larger than the corresponding up-quark GFFs of the kaon, while remaining smaller than the corresponding strange-quark GFFs of the kaon.
\begin{figure}[t]
\centering
\includegraphics[width=0.8\columnwidth]{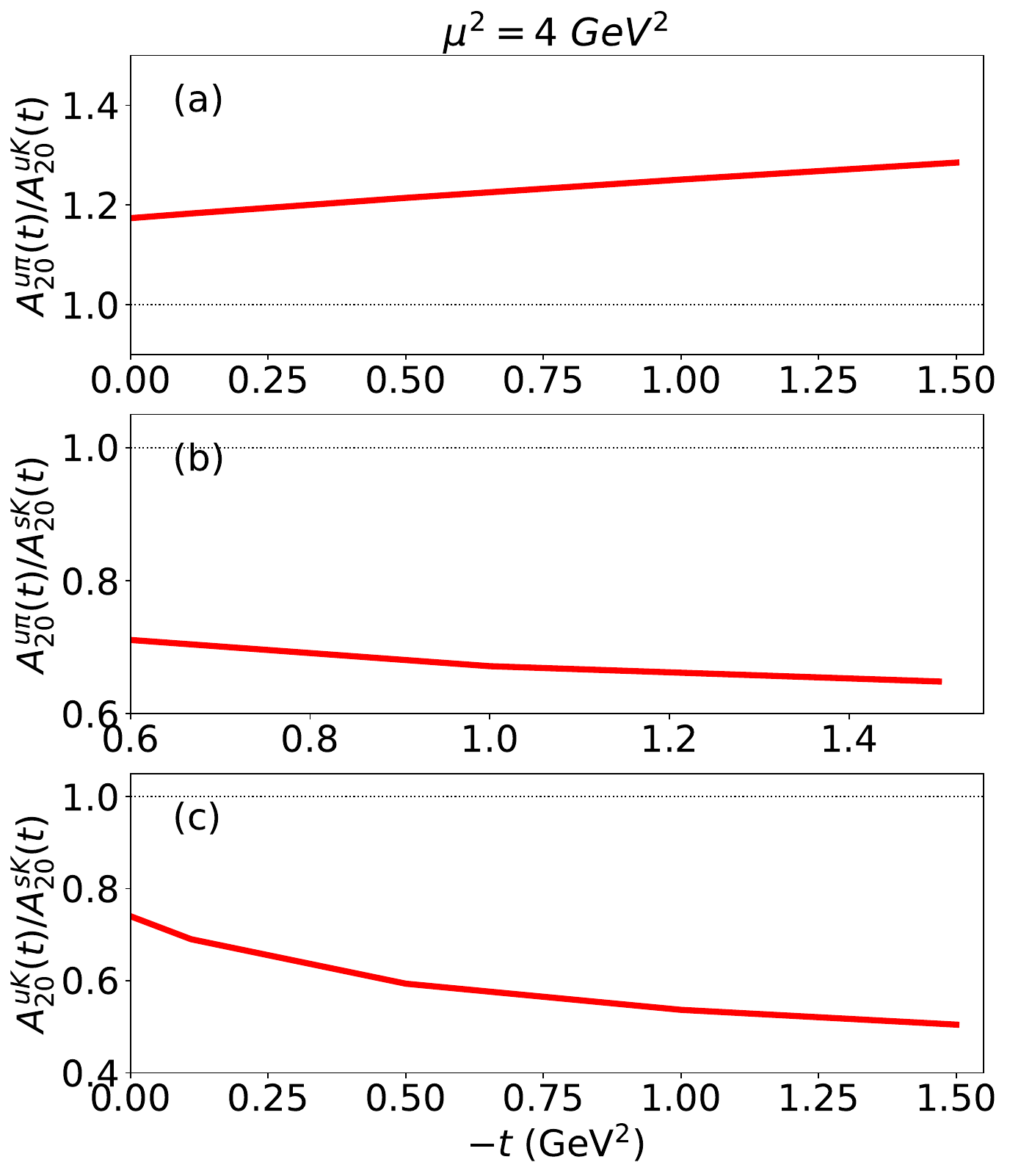}
\caption{\label{fig13} Ratio of $A_{20}^{u\pi}(t)/A_{20}^{qK}(t)$ at scales $\mu^2 =$ 4 GeV$^2$ with $q =(u,s)$.} 
\end{figure}

To clearly elucidate the differences between the pion and kaon GFFs, we calculate the ratios $A_{20}^{u\pi}(t)/A_{20}^{uK}(t)$, $A_{20}^{u\pi}(t)/A_{20}^{sK}(t)$, and $A_{20}^{uK}(t)/A_{20}^{sK}(t)$ at the scale $\mu^2=4~\mathrm{GeV}^2$ and compare them with the lattice QCD results of Ref.~\cite{Delmar:2024vxn}, as shown in Fig.~\ref{fig13}. In Fig.~\ref{fig13}(a), we find that our result for the ratio $A_{20}^{u\pi}(t)/A_{20}^{uK}(t)$ is consistent with the available lattice-QCD results~\cite{Delmar:2024vxn}, with the ratio remaining larger than unity. However, at larger values of $-t$, our result exhibits a larger magnitude than the lattice QCD result.

Figure~\ref{fig13}(b) shows the ratio $A_{20}^{u\pi}(t)/A_{20}^{sK}(t)$ as a function of $-t$. We find that this ratio decreases with increasing $-t$, in agreement with the lattice-QCD results~\cite{Delmar:2024vxn}. A similar behavior is observed for the ratio $A_{20}^{uK}(t)/A_{20}^{sK}(t)$ shown in Fig.~\ref{fig13}(c), which also decreases with increasing $-t$, as expected. This result is also consistent with the lattice-QCD result reported in Ref.~\cite{Delmar:2024vxn}.

\section{Summary and Conclusion}
\label{sec:sum}
In summary, we have investigated the $t$-dependence and nonzero skewness of the pion and kaon GPDs within the covariant NJL model using the proper-time regularization scheme, which regulates the quark loop integrals and provides an effective implementation of quark confinement. We have evaluated the pion and kaon valence-quark and gluon distributions for $-t=0$, $0.11$, $0.5$, $1.0$, and $1.5~\mathrm{GeV}^2$ and for skewness values $\xi=0.05$, $0.15$, and $0.25$ at the scales $\mu^2=4$ and $27~\mathrm{GeV}^2$. We have also computed the Mellin moments of the pion valence-quark distributions at the scales $\mu^2=4$ and $27~\mathrm{GeV}^2$.

We find that the pion and kaon valence-quark distributions exhibit only a weak dependence on $\xi$ at both scales, i.e., $\mu^2=4$ and $27~\mathrm{GeV}^2$. In contrast, they show a stronger dependence on $-t$, particularly in the regions $0\leq x\leq0.8$ for the kaon and $0\leq x\leq0.7$ for the pion. At both scales, the valence-quark distributions of the pion and kaon decrease with increasing $-t$. The gluon distributions likewise decrease with increasing $-t$, but exhibit only a weak dependence on both $\xi$ and $-t$ at $\mu^2=4$ and $27~\mathrm{GeV}^2$.

The generalized form factors of the pion and kaon at $\mu^2=4$ and $27~\mathrm{GeV}^2$ exhibit the following hierarchy:
\begin{eqnarray}
    A_{n0}^{uK}(t)<A_{n0}^{u\pi}(t)<A_{n0}^{sK}(t),
\qquad n=1,2,3,4.
\end{eqnarray}
Furthermore, we find that the ratio $A_{20}^{u\pi}(t)/A_{20}^{uK}(t)$ is consistent with the result from the lattice QCD simulation~\cite{Delmar:2024vxn}, with the ratio remaining larger than unity. However, at larger values of $-t$, our result exhibits a larger magnitude than the lattice QCD result. We further find that this ratio decreases with increasing $-t$, in agreement with the lattice-QCD results~\cite{Delmar:2024vxn}. A similar behavior is observed for the ratio $A_{20}^{uK}(t)/A_{20}^{sK}(t)$, which also decreases with increasing $-t$. Our result is also consistent with the lattice result in Ref.~\cite{Delmar:2024vxn}.

The results of this study are expected to provide useful theoretical input for the extraction of pion and kaon GFFs in future experiments at modern facilities, including the Electron-Ion Collider (EIC)~\cite{Arrington:2021biu}, the future Electron-Ion Collider in China (EicC)~\cite{Anderle:2021wcy}, the Apparatus for Meson and Baryon Experimental Research (AMBER)/COMPASS++ at CERN~\cite{Adams:2018pwt}, the J-PARC Hadron Experimental Facility Extension Project~\cite{Sakuma:2022twx,Aoki:2021cqa}, and the 22-GeV upgrade of Jefferson Lab (JLab)~\cite{Accardi:2023chb}, as well as future lattice-QCD simulations.

~

\section*{Acknowledgments}
This work was supported by the PUTI Q1 Research Grant from the University of Indonesia (UI) under Contract No. PKS-206/UN2.RST/HKP.05.00/2025 (F. C. and T. M.) and by the RCNP Collaboration Research Network program under Project No. COREnet 057 (P. T. P. H.). This work was also supported by the World Premier International Research Center Initiative for Sustainability with Knotted Chiral Meta Matter (WPI-SKCM$^2$) of Hiroshima University, MEXT, Japan.

\bibliographystyle{elsarticle-num}
\bibliography{main}
\end{document}